\documentclass{article}

\usepackage{amsmath, amssymb, parskip, physics, indentfirst, enumitem, comment, anyfontsize, multirow, array}

\usepackage[svgnames,dvipsnames,x11names]{xcolor}

\usepackage[linktocpage=true]{hyperref}

\usepackage{orcidlink}

\hypersetup{
colorlinks=true,
citecolor=darkblue,
linkcolor=reddish,
urlcolor=darkblue,
pdfauthor={},
pdftitle={},
pdfsubject={}
}

\definecolor{darkblue}{rgb}{0.15,0.35,0.55}
\definecolor{reddish}{rgb}{0.65, 0.2, 0.2}
\definecolor{darkgreen}{RGB}{50,150,0}
\definecolor{greyish2}{rgb}{.96,.96,.96}

\numberwithin{equation}{section}

\def\equationautorefname~#1\null{Equation #1\null}

\definecolor{turquoise}{RGB}{0,206,209}

\usepackage[normalem]{ulem}

\usepackage[nameinlink]{cleveref}

\DeclareFontFamily{OT1}{rsfs10}{}
\DeclareFontShape{OT1}{rsfs10}{m}{n}{ <-> rsfs10 }{}
\DeclareMathAlphabet{\mathscript}{OT1}{rsfs10}{m}{n}

\def\gsim{ \lower .75ex \hbox{$\sim$} \llap{\raise .27ex \hbox{$>$}} }
\def\lsim{ \lower .75ex \hbox{$\sim$} \llap{\raise .27ex \hbox{$<$}} }

\begin{document}
\renewcommand{\thefootnote}{\fnsymbol{footnote}}
\vspace{0truecm}
\thispagestyle{empty}

\vspace*{-0.3cm}

\begin{center}
{\fontsize{21}{18} \bf Ladder Symmetries of}\\[14pt]
{\fontsize{21}{18} \bf Higher Dimensional Black Holes}
\end{center}

\vspace{.15truecm}

\begin{center}
{\fontsize{13}{18}\selectfont
Roman Berens \orcidlink{0000-0003-1509-5463},${}^{\rm a}$\footnote{\texttt{roman.berens@vanderbilt.edu}} 
Lam Hui \orcidlink{0000-0001-7003-4132},${}^{\rm b}$\footnote{\texttt{lh399@columbia.edu}} 
Daniel McLoughlin \orcidlink{0009-0005-3535-2334},${}^{\rm b}$\footnote{\texttt{dcm2183@columbia.edu}} 
\\
  \vspace{.3cm}
Adam R. Solomon \orcidlink{0000-0002-7411-3951},${}^{\rm c, d}$\footnote{\texttt{adam.solomon@gmail.com}}
and 
John Staunton \orcidlink{0009-0004-1661-9577}${}^{\rm b}$\footnote{\texttt{j.staunton@columbia.edu}}
}
\end{center}
\vspace{.4truecm}

\centerline{{\it ${}^{\rm a}$Department of Physics \& Astronomy,
    Vanderbilt University, Nashville, TN 37212, U.S.A.}}
  
\vspace{.3cm}

\centerline{{\it ${}^{\rm b}$Center for Theoretical Physics, Department of Physics,}}
\centerline{{\it Columbia University, New York, NY 10027, U.S.A.}}

\vspace{.3cm}

\centerline{{\it ${}^{\rm c}$Department of Physics and Astronomy, McMaster University,}}
\centerline{{\it 1280 Main Street West, Hamilton ON, Canada}} 

\vspace{.3cm}

\centerline{{\it ${}^{\rm d}$Perimeter Institute for Theoretical Physics,}} 
\centerline{{\it 31 Caroline Street North, Waterloo ON, Canada}} 
 
\vspace{.25cm}

\vspace{.3cm}

\begin{abstract}

We compute the ladder operators for static tidal perturbations to higher-dimensional black holes. These operators map between solutions of the relevant equation of motion at different multipole orders. We focus on spin 0, 1, and 2 perturbations to the Schwarzschild--Tangherlini black hole and on spin 0 perturbations to the 5D Myers--Perry black hole. The ladder structure, used in conjunction with the existence of special ground state solutions, explains why the Love numbers of these higher-dimensional black holes vanish for specific combinations of the multipole moment and number of spacetime dimensions. This generalizes previous work on a ladder symmetry explanation for the vanishing of 4D black hole static Love numbers to higher dimensions. 

\end{abstract}

\newpage

\setcounter{tocdepth}{2}
\tableofcontents
\newpage
\renewcommand*{\thefootnote}{\arabic{footnote}}
\setcounter{footnote}{0}

\section{Introduction}
\label{Introduction}

Perturbations to black holes are more symmetric than one might naively expect, displaying many hidden symmetries. There has been recent progress in discovering and applying such hidden symmetries to explain the vanishing of black hole static tidal Love numbers, the parameters that encode the tidal response of a black hole to external static perturbations. These include symmetries in the near-zone approximation(s) \cite{Charalambous:2021a, Charalambous:2021b, Charalambous:2022, Hui:2022} as well as exact symmetries in the static regime, known as ladder symmetries \cite{Hui:2021, Berens:2022, BenAchour:2022, Sharma:2024, Rai:2024, Combaluzier-Szteinsznaider:2024sgb, Kehagias:2024, Gounis:2024, Berens:2025okm}. See also recent papers \cite{Parra-Martinez:2025, Lupsasca:2025}. The vanishing of static tidal Love numbers is a special property of static perturbations to black holes in 4D general relativity -- black hole Love numbers do not generically vanish when the perturbations are dynamical or if the theory contains higher-curvature corrections \cite{Kol:2011, Cardoso:2018, DeLuca:2022, Barbosa:2025,Caron-Huot:2025}. The situation for higher dimensional black holes within general relativity is similarly intriguing: some, but not all, Love numbers vanish. Our principal aim in this paper is to describe how the idea of ladder symmetries introduced by \cite{Hui:2021} can be used to understand this behavior.

The absence of a static tidal response for black holes is surprising from the point of view of boundary conditions. The computation of the static response of an object amounts to solving a second order differential equation, with two branches of solutions. At infinity, the two asymptotic behaviors of the relevant perturbation $\phi_\ell$ are growing ($\phi_{\ell}\sim r^{\ell}$) and decaying ($\phi_{\ell}\sim r^{-\ell-1}$) where $\ell$ is the angular momentum quantum number. Generically, one would expect to have an admixture of both behaviors. The Love number characterizing the tidal response (the decaying branch) of an object to an external perturbation (the growing branch) is the ratio of the decaying tail to the growing mode. Love numbers thus vanish when the decaying tail is absent. To determine this, one must consider the boundary condition at the surface of the object, which is the horizon for a black hole. At the horizon, the two asymptotic behaviors are constant (regularity) and divergent (irregularity). Imposing regularity at the horizon, as is appropriate for the black hole, turns out to imply --- after some computation --- that the solution is purely growing at infinity with no decaying tail. This is the vanishing Love number surprise: why does the generic expectation of an admixture of growing and decaying asymptotic behaviors fail? 

In four spacetime dimensions, this single-asymptote-behavior (for all $\ell$) can be explained by the ladder symmetries which arise from the representation theory of hidden SO(3,1) symmetries for static spin 0, 1, 2 perturbations around a Schwarzschild black hole \cite{Hui:2021, Berens:2022, Combaluzier-Szteinsznaider:2024sgb, Gray:2024, Berens:2025okm}\footnote{The vanishing of static Love numbers for a 4D Kerr black hole can be explained in a similar way, though it does not have the full SO(3,1) symmetry \cite{Hui:2021}.}. The SO(3,1) symmetries are geometric in origin and are a special property of four spacetime dimensions. Yet, the idea of ladder symmetries is still useful: the main goal of this paper is to show that the ladder symmetries have higher dimensional analogs and can be used to explain the special pattern of vanishing black hole Love numbers in higher dimensions. As is well known, and will be detailed below, not all black hole Love numbers vanish in higher dimensions, but some do, for special combinations of $\ell$ and the number of spacetime dimensions \cite{Kol:2011, Hui:2020, Charalambous:2023, Rodriguez:2023, Glazer:2024}. In all cases we consider in this paper (with a slight modification for 5D Myers--Perry), the following general pattern applies: that the equation of motion for the relevant perturbation $\phi_\ell$ can be cast in the form
\begin{equation}
\label{eq:EoMgeneralForm}
\left( D^+_{\hat \ell - 1} D^-_{\hat \ell} - \kappa_{\hat \ell}
  \right) \phi_{\hat \ell} = 0,
\end{equation}
where $D_{\hat{\ell}}^+$ and $D_{\hat{\ell}}^-$ are raising and lowering operators that map between solutions to the equations of motion at level $\hat{\ell}$ and level $\hat{\ell}\pm 1$, and $\kappa_{\hat{\ell}}$ is an $\hat{\ell}$-dependent constant. Here, $\hat \ell \equiv \ell / (D-3)$ where $D$ is the number of spacetime dimensions. It's important that $D^{\pm}_{\hat \ell}$ are first order differential operators. If there exists a value of $\hat{\ell}$ for which $\kappa_{\hat{\ell}} = 0$, then the equation of motion factorizes into $D^+_{\hat \ell - 1} D^-_{\hat \ell}\phi_{\hat \ell} = 0$. We refer to this special value of $\hat{\ell}$ as a \textit{ground level}. In that case, the natural guess for the solution is $D^-_{\hat \ell} \phi_{\hat \ell} = 0$ (by analogy with the quantum harmonic oscillator whose ground state is annihilated by the lowering operator). We will refer to a solution of the equation of motion that satisfies a first-order differential equation of this form as a \textit{ground state} solution. The fact that $D^-_{\hat \ell} \phi_{\hat{\ell}} = 0$ means it is no longer a surprise that a single asymptote at the horizon connects to a single asymptote at infinity. It is then a matter of checking that the ground state asymptotes to be regular at the horizon and to a purely growing mode at infinity. As we will see, the structure of the raising operator $D_{\hat{\ell}}^+$ tells us that the higher $\hat \ell$ solutions, constructed by raising successively from the ground state, will share the properties of regularity at the horizon and purely growing at infinity. It is important that this argument only applies to those solutions with an $\hat \ell$ that is reachable by an integer number of steps from the ground level. This will explain the special pattern of vanishing black hole Love numbers in higher dimensions. 

We note that defining Love numbers as the ratios of the decaying and growing amplitudes at infinity invites the valid criticism that this is naively coordinate dependent and therefore potentially misleading. A good way to address this problem is to define the Love numbers via the worldline point particle effective field theory (EFT), following \cite{Goldberger:2004jt}. The analyses of \cite{Hui:2020, Glazer:2024} show that our naive definition used in this paper matches the more careful EFT definition for the higher dimensional black holes of interest.

In this paper, we find a symmetry explanation for the specific tuning of multipole moment and spacetime dimension for which linear static Love numbers of higher dimensional black holes vanish. We quote the equations of motion for perturbations to both Schwarzschild--Tangherlini and 5D Myers--Perry from other sources and give a brief re-derivation of these results in appendices. Our new results are as follows. We find the ladder operators for these higher-dimensional black holes, which raise and lower by integer values of $\hat{\ell}$. The question of whether Love numbers vanish can then be traced, through climbing up and down the ladder, to the existence of a ground state, i.e. that there is a particular choice of $\hat{\ell}$ (a ground level) for which $\kappa_{\hat{\ell}}$ in \autoref{eq:EoMgeneralForm} is equal to zero. As we will show, this only occurs for those specific combinations of multipole moment and spacetime dimension at which the Love numbers vanish. This will further make manifest the subtle combination of parameters for which the scalar Love numbers of the 5D Myers--Perry black hole vanish. In addition, we will connect our analysis to the horizontal symmetries of \cite{Hui:2021, BenAchour:2022} and show how these generalize to higher dimensions.

\textit{Conventions and Structure:} We will use the mostly plus signature and work in natural units where $c = \hbar = G = 1$. Unless otherwise stated, we work in $D$ spacetime dimensions. In \autoref{s:Spin0} -- \autoref{s:Spin2}, we compute the ladder structures and the multipole moments at which static Love numbers vanish in $D$-dimensional Schwarzschild--Tangherlini. In \autoref{s:MyersPerry}, we apply the same techniques to spin 0 perturbations to the 5D Myers--Perry black hole. Many technical details as well as a discussion on alternative symmetry perspectives are contained in the appendices. 

\section{Spin 0}
\label{s:Spin0}
In this section, we will study a static massless spin 0 particle living in a Schwarzschild--Tangherlini background, which is described by the line element \cite{Tangherlini:1963}
\begin{equation}
\label{eq:STMetric}
\dd{s}^2 = -f\left(r\right) \dd{t}^2 + \frac{1}{f\left(r\right)} \dd{r}^2 + r^2 \dd{\Omega}^2,
\end{equation}
where $\dd{\Omega}^2$ is the metric on the $\left(D-2\right)$-sphere and
\begin{equation}
f\left(r\right) \equiv 1 - \left(\frac{r_s}{r}\right)^{D - 3}.
\end{equation}
As we will see, the ladder structure for this particle, upon the identification of the angular momentum $\ell$ in four dimensions with $\hat{\ell} = \ell/(D - 3)$ in $D$ dimensions, is exactly the same as that of the 4D spin 0 particle in 4D Schwarzschild \cite{Hui:2021}. This section therefore doubles as a pedagogical review of the ladder symmetry arguments. 

\subsection{Ladder Structure}
To begin, the action for a massless, minimally coupled scalar, $\phi$, in $D$ spacetime dimensions is
\begin{equation}
S = -\frac{1}{2} \int \partial_\mu \phi \partial^\mu \phi \sqrt{-g} \dd[D]{x}.
\end{equation}
Since we are interested in the static and conservative response of the black hole, we set time derivatives to zero, yielding the following equation of motion:
\begin{equation}
\label{eq:Spin0ScalarEom}
\frac{1}{r^D} \partial_r\left(r^{D - 2} f\left(r\right)\partial_r \phi\right) + \nabla^2_{S^{D - 2}}\phi = 0,
\end{equation}
where $\nabla^2_{S^{D - 2}}$ is the Laplacian on the $\left(D-2\right)$-sphere.

We use the rotational invariance of the background to decompose the scalar into spherical harmonics as
\begin{equation}
\phi \equiv \sum_{\ell, M}\phi_\ell\left(r\right) Y_\ell^{M}\left(\theta\right),
\end{equation}
where the $Y_\ell^{M}\left(\theta\right)$ are hyperspherical harmonics\footnote{Note that the many magnetic quantum number indices have been grouped into the same multi-index $M$ since we will only be interested in the $\ell$-dependence of the $Y_\ell^{\ M}$.} defined on the $\left(D - 2\right)$-sphere, with the spherical Laplacian eigenvalue equation
\begin{subequations}
\begin{align}
\nabla_{S^{D-2}}^{2}Y_\ell^{M} &=  -\ell\left(\ell + D - 3\right)Y_\ell^{M}, \\
&= -\left(D-3\right)^{2}\hat{\ell}\left(\hat{\ell} + 1\right)Y_\ell^{M},
\end{align}
\end{subequations}
and where we introduce $\hat{\ell}$,
\begin{equation}
\label{eq:lhat}
\hat{\ell} \equiv \frac{\ell}{D-3},
\end{equation}
so that the eigenvalue is most similar to the $D = 4$ case. For a review of relevant information about hyperspherical harmonics, see \cite[Appendix A]{Hui:2020}. From this point forward, it will be helpful to work with $\hat{\ell}$ (which need not be an integer) instead of $\ell$. In particular, we will index the radial part of the field with $\hat{\ell}$ although the hyperspherical harmonics are still indexed by the integer $\ell$. 

It is helpful to employ a variable change that maps the equation of motion onto one that is equivalent to a minimally coupled and static scalar in $D = 4$ dimensions. Namely, we define
\begin{equation}
\label{eq:ydef}
y \equiv \left(\frac{r}{r_s}\right)^{D - 3} \qquad \text{and} \qquad \Delta \equiv y\left(y - 1\right).
\end{equation}
In terms of $\hat{\ell}$ and $y$, the equation of motion takes the form
\begin{equation}
\label{eq:Spin0EOM}
\partial_y\left(\Delta \partial_y \phi_{\hat{\ell}}\left(y\right)\right) - \hat{\ell}\left(\hat{\ell} + 1\right)\phi_{\hat{\ell}}\left(y\right) = 0.
\end{equation}
As promised, the four-dimensional equation of motion for a static massless scalar (see \cite[eq. (2.2)]{Hui:2021}) maps onto this equation of motion with the identification $\ell \mapsto \hat{\ell}$, $r \mapsto y$, and $r_s \mapsto 1$. Then, the authors of \cite{Hui:2021} find a 4D ladder structure that connects a solution of the equation of motion at level $\ell$ to one at level $\ell\pm 1$, which in our higher dimensional case becomes a ladder in $\hat{\ell}$:
\begin{subequations}
\begin{align}
\label{eq:RaisingOpScalar}
D_{\hat{\ell}}^+ &= -\Delta \partial_y - \frac{\hat{\ell} + 1}{2}\Delta', \\
D_{\hat{\ell}}^- &= \Delta \partial_y - \frac{\hat{\ell}}{2} \Delta'.
\label{eq:LoweringOpScalar}
\end{align}
\end{subequations}
The computation of ladder operators is most easily accomplished with the help of hypergeometric identities that are reviewed in \autoref{a:HypergeometricIdentities}.

The interpretation of $D^{\pm}_{\hat{\ell}}$ as ladder operators in $\hat{\ell}$ is made concrete by defining the ``Hamiltonian" as $-\Delta$ times the equation of motion:
\begin{equation}
H_{\hat{\ell}} = -\Delta\left[\partial_y\left(\Delta \partial_y \right) - \hat{\ell}\left(\hat{\ell} + 1\right)\right].
\end{equation}
In terms of the ladder operators, the Hamiltonian is expressible as
\begin{subequations}
\begin{align}
\label{eq:HwithDminus}
H_{\hat{\ell}} &= D_{\hat{\ell} - 1}^+ D_{\hat{\ell}}^- - \frac{\hat{\ell}^{\: 2}}{4},\\
&= D_{\hat{\ell} + 1}^- D_{\hat{\ell}}^+ - \frac{\left(\hat{\ell} + 1\right)^2}{4}.\label{eq:HwithDplus}
\end{align}
\end{subequations}
Relatedly, the intertwining relation among the ladder operators is
\begin{equation}
D_{\hat{\ell} + 1}^- D_{\hat{\ell}}^+ - D_{\hat{\ell} - 1}^+ D_{\hat{\ell}}^- = \frac{2\hat{\ell} + 1}{4}.
\end{equation}
Finally, the operator algebra between the Hamiltonian and the ladder operators that justifies their interpretation as such is 
\begin{subequations}
\label{eq:LadderOperatorAlgebra}
\begin{align}
H_{\hat{\ell} + 1} D_{\hat{\ell}}^+ &= D_{\hat{\ell}}^+ H_{\hat{\ell}}, \\ 
H_{\hat{\ell} - 1} D_{\hat{\ell}}^- &= D_{\hat{\ell}}^- H_{\hat{\ell}}.
\end{align}
\end{subequations}
Note that, although we derived \autoref{eq:RaisingOpScalar} and \autoref{eq:LoweringOpScalar} using the hypergeometric function identities approach outlined in \autoref{a:HypergeometricIdentities}, which makes use of the full solutions of the equation of motion, this is not necessary. The same operators can be bootstrapped from their well-defined algebra with the Hamiltonian, \autoref{eq:LadderOperatorAlgebra}.

\subsection{The Vanishing of Love Numbers}
\label{ss:TheVanishingofLoveNumbers}

We claim that the ladder symmetry explains when the static Love numbers vanish for any spacetime dimension. The ladder operators of \autoref{eq:RaisingOpScalar}--\autoref{eq:LoweringOpScalar} raise and lower $\hat{\ell}$ by one, regardless of whether or not $\hat{\ell}$ is an integer. As explained in \autoref{Introduction}, if there exists a choice of $\hat{\ell}$ (a ground level) for which the second-order equation of motion can be reduced to a first-order one, i.e., for which there exists a ground state, then the two branches of solutions to the equation of motion completely decouple from one another, each solving a different first-order differential equation. The branch we care about is the one that is regular at the horizon. If we find that the ground state of this branch is also purely growing at infinity, then the Love number for the ground state vanishes. By explicit computation, we can then check that the boundary conditions at the horizon and infinity are preserved by the ladder operators. In this case, repeatedly acting with the raising operator on a ground state with vanishing Love number guarantees that the Love numbers vanish for all $\hat{\ell}$ connected to the ground state by an integer.

Inspection of the Hamiltonian in \autoref{eq:HwithDminus} reveals that the ground level for spin 0 perturbations is $\hat{\ell} = 0$. At this level, \autoref{eq:HwithDminus} reduces to
\begin{equation}
H_{0}\phi_{0} = D_{-1}^{+}D_{0}^{-}\phi_{0} = 0.
\end{equation}
Any solution of
\begin{equation}
\label{eq:ScalarGroundStateMinus}
D_{0}^{-}\phi_{0}=\Delta\partial_{y}\phi_{0}=0,
\end{equation}
which we refer to as a ``ground state'', is guaranteed to also satisfy $D_{-1}^{+}D_{0}^{-}\phi_{0} = 0$, thereby reducing the second-order differential equation to a first-order one. In this specific case, by inspection, $\phi_{0}$ is a constant. This solution is regular at the horizon and has no decaying tail at infinity. Furthermore, acting repeatedly with $D_{\hat{\ell}}^{+}$, defined in \autoref{eq:RaisingOpScalar}, will only increase the power of $y$ each time:
\begin{equation}
\phi_{\hat{\ell}} \propto D_{\hat{\ell} - 1}^{+}\dots D_{0}^{+}\phi_{0}.
\end{equation}
This branch of solutions will therefore only ever be purely growing at infinity.

To conclude the argument, we recall that the vanishing of the Love numbers requires both a ladder structure and the existence of a ground state. Since the ladder increases or decreases the value of $\hat{\ell}$ by 1, we conclude that the ground state only exists for $\hat{\ell}$ being an integer. The Love numbers therefore vanish for all integer $\hat{\ell}$, in agreement with the conclusion in e.g. \cite{Kol:2011, Hui:2020}.

An important side note is that although we have discussed a ladder structure, this can be rephrased as a symmetry. One type of symmetry is a \textit{vertical ladder symmetry} that mixes an $\hat{\ell}$ and $\hat{\ell} - 1$ level \cite{Hui:2021}. The vertical ladder symmetry acts on the fields as
\begin{equation}
\label{verticalSymm}
\delta \phi_{\hat{\ell}} = D_{\hat{\ell} - 1}^+ \phi_{\hat{\ell} - 1} \qquad \text{and} \qquad \delta \phi_{\hat{\ell} - 1} = -D_{\hat{\ell}}^- \phi_{\hat{\ell}}.
\end{equation}
This symmetry encodes the ladder structure\footnote{For the case of spin 0, an alternative way to see this symmetry is to introduce the operators $A_{\hat{\ell}} = D_{\hat{\ell}}^-$ and $A_{\hat{\ell}}^\dagger = D_{\hat{\ell} - 1}^+$. After an appropriate redefinition, the Hamiltonian can be interpreted, at level $\hat{\ell}$, as the supersymmetric partner of the Hamiltonian at level $\hat{\ell} + 1$. The story is identical to that presented in \cite[Appendix E]{Hui:2021}.}. A second type of symmetry is a \textit{horizontal ladder symmetry}, which we discuss in more detail in \autoref{a:Horizontal}.

\section{Spin 1}
In this section, we turn our attention to a massless spin 1 particle also living in a Schwarzschild--Tangherlini background \autoref{eq:STMetric}. Due to the spherical symmetry of the background, the spin-1 field $A_{\mu}$ can be decomposed into two master variables: one scalar variable, $\Psi^S$, and one vector variable, $\Psi^V$. For a review of the decomposition of gauge fields and the derivation of the equations of motion for $\Psi^S$ and $\Psi^V$, see \autoref{a:HarmonicDecompositions}, which follows \cite{Hui:2020}.

\subsection{Vector Perturbations}
\label{s:OddParityDegreeOfFreedom}
We begin with the equation of motion for the vector master variable, \autoref{eq:Spin1OddEoM}:
\begin{equation}
\label{eq:spin1VectorEom}
f\dv[2]{\Psi^V}{r} + f'\dv{\Psi^V}{r} - \left(\frac{\left(\ell + 1\right)\left(\ell + D - 4\right)}{r^2} + \frac{\left(D - 4\right)\left[\left(D - 6\right) f + 2r f'\right]}{4r^2}\right) \Psi^V = 0.
\end{equation}
As in the scalar case, it is beneficial to transform to $\hat{\ell}$, defined in \autoref{eq:lhat}, and to the dimensionless coordinate $y$, defined in \autoref{eq:ydef}. We further perform the following field redefinition:
\begin{equation}
\label{eq:FieldReDef}
\Psi^V\left(y\right) = y^{-\frac{D - 4}{2\left(D - 3\right)}} \phi^V\left(y\right).
\end{equation}
This transforms the equation of motion into a form for which the $D \rightarrow 4$ limit is particularly simple:
\begin{equation}
\label{eq:Spin1VectorEomPhi}
\left[\partial_y\left(\Delta \partial_y\right) - 2\left(y - 1\right)\partial_y + \frac{\mu^2 - 1}{y} - \hat{\ell}\left(\hat{\ell} + 1\right)\right]\phi^V_{\hphantom{V}\hat{\ell}}\left(y\right) = 0,
\end{equation}
where
\begin{equation}
\label{eq:muDefinition}
\mu \equiv \frac{1}{D - 3}
\end{equation}
contains the dimension dependence of the equation of motion. The $D \rightarrow 4$ limit is then found by the replacements $y \mapsto r/r_s$, $\hat{\ell} \mapsto \ell$, and $\mu \mapsto 1$.

Utilizing the procedure outlined in \autoref{a:HypergeometricIdentities}, the ladder operators, which connect solutions at different values of $\hat{\ell}$, are
\begin{subequations}
\begin{align}
\label{eq:RaisingSpin1Vector}
D^+_{\hat{\ell}} &\equiv -\Delta \partial_y - \frac{\hat{\ell}^{\:2} \Delta' + 2\hat{\ell} y - \mu^2 + 1}{2\left(\hat{\ell} + 1\right)},\\
D^-_{\hat{\ell}} &\equiv \Delta\partial_y - \frac{\hat{\ell}^{\:2} \Delta' + 2\hat{\ell}\left(y - 1\right) - \mu^2}{2\hat{\ell}}.
\label{eq:LoweringSpin1Vector}
\end{align}
\end{subequations}
This can be verified by their algebra with the Hamiltonian, defined to be $-\Delta$ times the equation of motion:
\begin{equation}
\label{eq:HamiltonianSpin1Odd}
H_{\hat{\ell}} \equiv -\Delta\left[\partial_y\left(\Delta \partial_y\right) - 2\left(y - 1\right)\partial_y + \frac{\mu^2 - 1}{y} - \hat{\ell}\left(\hat{\ell} + 1\right)\right].
\end{equation}
Written in terms of ladder operators, we find
\begin{subequations}
\begin{align}
\label{eq:Spin1VectorHamiltonian}
H_{\hat{\ell}} &= D^+_{\hat{\ell} - 1} D^-_{\hat{\ell}} - \frac{\left(\hat{\ell}^{\:2} - \mu^2\right)^2}{2\hat{\ell}^{\:2}},\\
&= D^-_{\hat{\ell} + 1} D^+_{\hat{\ell}} - \left(\frac{\left(\hat{\ell} + 1\right)^2 - \mu^2}{2\left(\hat{\ell} + 1\right)}\right)^2.
\end{align}
\end{subequations}
In addition, the intertwining relation is
\begin{equation}
D_{\hat{\ell} + 1}^{-} D_{\hat{\ell}}^{+} -D_{\hat{\ell} - 1}^{+} D_{\hat{\ell}}^{-} = \frac{2\hat{\ell} + 1}{4}\left(1 - \frac{\mu^{4}}{\hat{\ell}^{\:2}\left(\hat{\ell}^{\:2} + 1\right)}\right).
\end{equation}
With these definitions, the same ladder operator algebra as \autoref{eq:LadderOperatorAlgebra} is obeyed.

From \autoref{eq:Spin1VectorHamiltonian}, we see that the ground level is $\hat{\ell} = \pm \mu$. This simplifies the differential equation to
\begin{equation}
\label{eq:Spin1OddReducedEoM}
H_{\pm \mu} \phi^V_{\hphantom{V} \pm\mu} = D_{\pm\mu - 1}^+ D_{\pm \mu}^- \phi^V_{\hphantom{V} \pm \mu} = 0.
\end{equation}
It follows that we can choose a ground state solution that obeys the first-order equation
\begin{equation}
D_{\pm \mu}^- \phi^V_{\hphantom{V} \pm \mu} = 0,
\end{equation}
with the solution
\begin{equation}
\label{eq:Spin1VReg}
\phi^V_{\hphantom{V}\pm \mu}\left(y\right) \propto y^{\pm \mu + 1}.
\end{equation}
This solution is regular at the horizon, $y = 1$. Higher level-$\hat{\ell}$ growing modes are found by acting with a chain of raising operators as occurred in the scalar case, and direct computation reveals that this does not mix between the solution branches. We then conclude that the Love numbers vanish for the branch of solutions connected to this ground state $\phi^V_{\hphantom{V}\hat{\ell}}$ by an integer. 

If we climb the ladder for $\hat{\ell} = \pm \mu$, we find that the Love numbers vanish for any $\hat{\ell}\geq 0$ satisfying 
\begin{equation}
\label{eq:hatlSpin1V}
\hat{\ell} = n \pm \frac{1}{D - 3}, \qquad n \in \mathbb{Z}_{\geq 0}.
\end{equation}
For example, the Love numbers vanish for all values of $\hat{\ell}$ in $D = 4$ and for half-integer $\hat{\ell}$ in $D = 5$. Note that our analysis predicts two ground levels from which we can build a ladder of solutions with vanishing Love numbers: a ground level that starts at positive $\ell$ and one that starts at negative $\ell$. Negative values of $\ell$ are clearly unphysical since $\ell$ indexes spherical harmonics. However, since the ladder operators preserve the boundary conditions at the horizon and infinity, it is possible to start with an unphysical solution at negative $\ell$ and climb the ladder to a physical one at positive $\ell$. This physical solution is a perfectly good solution of the equation of motion with the desired regularity at the horizon and purely growing behavior at infinity, regardless of how we arrived at it. In $D = 4$ and $D=5$, these two branches are degenerate for physical values of $\hat{\ell}$, yet for $D>5$ this yields two distinct and non-overlapping branches of solutions. This is in agreement with the result in \cite[eq. 4.45]{Hui:2020}. We finally note that for spin 1 fields, tidal perturbations start at $\ell=1$. We therefore find vanishing Love numbers for those values of $\hat{\ell}$ in \autoref{eq:hatlSpin1V} with $\hat{\ell}>0$.

\subsection{Scalar Perturbations}
\label{s:EvenParityDegreeOfFreedom}

The equation of motion for the scalar master variable, \autoref{eq:Spin1EvenEoM}, is
\begin{equation}
\label{eq:Spin1ScalarEom}
f \dv[2]{\Psi^S}{r} + f' \dv{\Psi^S}{r} - \left(\frac{\ell\left(\ell + D - 3\right)}{r^2} + \frac{\left(D - 4\right)\left[\left(D - 2\right) f - 2r f'\right]}{4r^2}\right) \Psi^S = 0.
\end{equation}
Once again transforming to $\hat{\ell}$ and $y$, and imposing the field redefinition 
\begin{equation}
\Psi^S\left(y\right) = y^{-\frac{D - 4}{2\left(D - 3\right)}}\phi^S\left(y\right),
\end{equation}
the equation of motion takes the simple form
\begin{equation}
\label{eq:Spin1ScalarEomPhi}
\left[\partial_y\left(\Delta \partial_y\right) - 2\left(y - 1 \right)\partial_y - \hat{\ell}\left(\hat{\ell} + 1\right)\right]\phi^S_{\hphantom{S}\hat{\ell}} = 0.
\end{equation}
The $D \rightarrow4 $ limit is then straightforwardly found by the replacements $y \mapsto r/r_{s}$ and $\hat{\ell} \mapsto \ell$. This form of the equation of motion also makes the electromagnetic duality of spin 1 in $D = 4$ spacetime dimensions manifest \cite{Hui:2020,Solomon:2023ltn}: it coincides with \autoref{eq:Spin1VectorEomPhi} when $\mu = 1$, i.e., when $D = 4$. 

For this equation of motion we can once again find raising and lowering operators
\begin{subequations}
\begin{align}
D_{\hat{\ell}}^+ &= -\Delta \partial_y - \frac{\hat{\ell}^{\:2} \Delta' + 2\hat{\ell} y}{2\left(\hat{\ell} + 1\right)},\\
D_{\hat{\ell}}^- &= \Delta\partial_y - \frac{\hat{\ell}^{\:2} \Delta' + 2\hat{\ell}\left(y - 1\right) - 1}{2\hat{\ell}}.
\end{align}
\end{subequations}
If we again define a Hamiltonian by multiplying the equation of motion by $-\Delta$,
\begin{equation}
\label{eq:HamiltonianSpin1Even}
H_{\hat{\ell}} \equiv -\Delta\left[\partial_y\left(\Delta \partial_y\right) - 2\left(y - 1\right)\partial_y - \hat{\ell}\left(\hat{\ell} + 1\right)\right],
\end{equation}
we can express it in terms of the ladder operators:
\begin{subequations}
\begin{align}
\label{eq:Spin1ScalarHamiltonian}
H_{\hat{\ell}} &= D_{\hat{\ell} - 1}^+ D_{\hat{\ell}}^- - \left(\frac{\hat{\ell}^{\:2} - 1}{2\hat{\ell}}\right)^2,\\
&= D_{\hat{\ell} + 1}^- D_{\hat{\ell}}^+ - \left(\frac{\left(\hat{\ell} + 1\right)^{\:2} - 1}{2\left(\hat{\ell} + 1\right)}\right)^2.
\end{align}
\end{subequations}
The intertwining relation is 
\begin{equation}
D_{\hat{\ell} + 1}^- D_{\hat{\ell}}^+ - D_{\hat{\ell} - 1}^+ D_{\hat{\ell}}^- = \frac{2\hat{\ell} + 1}{4}\left(1 - \frac{1}{\left(\hat{\ell}\left(\hat{\ell} + 1\right)\right)^2}\right).
\end{equation}
With these definitions, the same ladder operator algebra as \autoref{eq:LadderOperatorAlgebra} is obeyed.

We now run the ground-state argument again. By inspection of \autoref{eq:Spin1ScalarHamiltonian} the ground level is $\hat{\ell} = 1$, at which the Hamiltonian reduces to
\begin{equation}
H_{1} \phi^S_{\hphantom{S} 1} = D_{0}^+ D_{1}^- \phi^S_{\hphantom{S}1} = 0.
\end{equation}
Thus, we find the first-order differential equation
\begin{equation}
D_1^- \phi^S_{\hphantom{S}1} = 0,
\end{equation}
which is solved by
\begin{equation}
\phi^S_{\hphantom{S}1}\left(y\right) \propto y^{2}.
\end{equation}
This is regular at the horizon ($y = 1$) and, as in the spin 0 case, acting repeatedly with the raising operator yields a purely growing mode with no decaying tail at infinity. The entire family of growing modes at all $\hat{\ell}$ can then be generated from the $\hat{\ell} = 1$ mode\footnote{Note one could have also attempted to start at the $\hat{\ell} = -1$ ground level. However, for physical values of $\ell\geq0$, the solutions generated starting at $\hat{\ell}=-1$ overlap with those starting at $\hat{\ell} = 1$ as the ladders raise by 1 each time. We therefore discard the negative ground level.}. We therefore reach a similar conclusion: the spin 1 scalar Love numbers vanish for all non-zero integer $\hat{\ell}$ (again, only those $\ell\geq1$ are spin 1 tidal perturbations). This agrees with the result found in \cite[eq. (4.35)]{Hui:2020}.

\section{Spin 2}
\label{s:Spin2}
We now consider spin 2 perturbations on the Schwarzschild--Tangherlini background. The degrees of freedom of a generic spin 2 perturbation, $h_{\mu\nu}$, upon harmonic decomposition, reduce to three master variables: a tensor variable, $\Psi^T$, a vector variable, $\Psi^{\text{RW}}$ (which coincides with the Regge-Wheeler variable in four dimensions \cite{Regge:1957}), and a scalar variable, $Z$ (which is related to the Zerilli variable in four dimensions \cite{PhysRevLett.24.737}). A review of the harmonic decomposition of the metric and subsequent derivation of the equations of motion for these variables, done in \cite{Hui:2020}, is found in \autoref{a:HarmonicDecompositions}.

\subsection{Tensor Perturbations}
\label{s:TensorPerturbations}
The equation of motion for the tensor perturbations, \autoref{eq:Spin2TensorEoM}, is 
\begin{equation}
f\dv[2]{\Psi^T}{r} + f' \dv{\Psi^T}{r} - \left(\frac{\ell\left(\ell + D - 3\right) + 2\left(D - 3\right)}{r^2} + f'\frac{D - 6}{2r} + f\frac{D\left(D - 14\right) + 32}{4r^2}\right)\Psi^T = 0.
\end{equation}
As was remarked in \cite{Hui:2020}, direct substitution of $f$ reveals that this is identical to the equation of motion for the spin 0 scalar perturbations in \autoref{eq:Spin0ScalarEom}. The ladder structure is therefore identical to the spin 0 scalar case and the Love numbers vanish for all integer $\hat{\ell}$. However, a key difference between the spin 0 and spin 2 cases is that the only spin 2 tidal perturbations are those for $\ell\geq2$ \cite{Martel:2005}. Indeed, the tensor spherical harmonics used in the harmonic decomposition of the metric perturbation are only defined for $\ell\geq2$.

\subsection{Vector Perturbations}
\label{s:VectorPerturbations}
The equation of motion for the generalized Regge-Wheeler variable, \autoref{eq:Spin2VectorEoM}, is
\begin{equation}
f\dv[2]{\Psi^{\text{RW}}}{r} + f' \dv{\Psi^{\text{RW}}}{r} - \left(\frac{\left(\ell + 1\right)\left(D - 4 + \ell\right)}{r^2} + f\frac{\left(D - 4\right)\left(D - 6\right)}{4r^2} - f' \frac{D + 2}{2r}\right) \Psi^{\text{RW}} = 0.
\end{equation}
Transforming $r$ to $y$ and $\ell$ to $\hat{\ell}$ as before and effecting the same field transformation as \autoref{eq:FieldReDef}, we find the simplified equation of motion
\begin{equation}
\left[\partial_y\left(\Delta \partial_y\right) - 2\left(y - 1\right)\partial_y + \frac{\nu^2 - 1}{y} - \hat{\ell}\left(\hat{\ell} + 1\right)\right]\phi^{\text{RW}}_{\hphantom{RW}\hat{\ell}}\left(y\right) = 0,
\end{equation}
where
\begin{equation}
\label{eq:nuDefinition}
\nu \equiv \frac{D - 2}{D - 3}.
\end{equation}
Interestingly, this is the same equation of motion as in the case of the spin 1 vector modes, equation \autoref{eq:Spin1VectorEomPhi}, if we identify $\mu$ and $\nu$. In reality, these values cannot coincide; however, this observation does make the computation of the ladder operators immediate: we simply take the spin 1 vector results and replace $\mu \rightarrow \nu$. The ladder operators are
\begin{subequations}
\begin{align}
\label{eq:Spin2VectorLadders}
D_{\hat{\ell}}^+ &= -\Delta\partial_y - \frac{\hat{\ell}^{\:2}\Delta' + 2\hat{\ell} y - \nu^2 + 1}{2\left(\hat{\ell} + 1\right)},\\
D_{\hat{\ell}}^- &= \Delta\partial_y - \frac{\hat{\ell}^{\:2} \Delta' - 2\hat{\ell} \left(y - 1\right) - \nu^2}{2\hat{\ell}}. 
\end{align}
\end{subequations}
Note that for $D=4$ these reduce to the Regge-Wheeler ladders of \cite{Hui:2021}. Since the ladder structure is the same, with the replacement of $\mu$ with $\nu$, we find that the ground level is $\hat{\ell} = \pm \nu$. Climbing the ladder then increases $\hat{\ell}$ by an integer $n$ and so we find that the Love numbers vanish for
\begin{equation}
\label{eq:LhatConditionSpin2Vector}
\hat{\ell} = n \pm \frac{D - 2}{D - 3}, \qquad n \in \mathbb{Z}_{\geq 0}.
\end{equation}

Our conclusion is similar to the spin 1 vector case: two ground levels exist from which one can build solutions to physical values of $\ell$, even if the starting point of the ladder is an unphysical $\ell<0$. Once again, for $D = 4$, Love numbers vanish for all integer $\ell$, and for $D = 5$, they vanish for all half-integer $\hat{\ell}$. This coincides\footnote{Note that in \cite{Hui:2020} the condition for the spin 2 vector Love number vanishing is given as $\hat{\ell} = m\pm\frac{1}{D-3}$ for integer $m$. This is equivalent to our condition in \autoref{eq:LhatConditionSpin2Vector} since we can re-write $\hat{\ell} = n \pm \frac{D - 2}{D - 3} = m \pm \frac{1}{D - 3}$ where $m = n \pm 1$ is also an integer.} with the result in \cite[eq. (4.56)]{Hui:2020}. We note that in $D = 4$, there is only one branch of solutions starting at $\ell = 2$ as the solutions built from the negative ground level degenerate with those built from the positive ground level for physical $\ell$. Additionally, the only tidal perturbations are those for $\ell\geq2$\footnote{This does not mean, however, that there are no $\ell = 0$ and $\ell = 1$ modes. Instead, these modes have the interpretation of mass ($\ell = 0$) and spin ($\ell = 1$) perturbations as opposed to static responses to an external perturbation. See a review \href{https://www.its.caltech.edu/~kip/PubScans/II-175.pdf}{here} by Thorne and the results of \cite{Martel:2005}.}. 

\subsection{Scalar Perturbations}
\label{ss:Spin2ScalarPerturbations}
Finally, there is the scalar master variable $Z$, the higher-dimensional generalization of the Zerilli variable. The equation of motion, \autoref{eq:Spin2ScalarEoM}, is
\begin{equation}
\begin{aligned}
f\dv[2]{Z}{r} + \left(\frac{D - 6}{D - 3} f' + \frac{D}{r}\right)\dv{Z}{r} - \left(\frac{D - 4}{\left(D - 3\right)\left(D - 2\right)} f'' + \frac{\left(\ell - 1\right)\left(d + \ell - 2\right)}{r^2}\right) Z = 0.
\end{aligned}
\end{equation}
As before, we can use the variables $y$ and $\hat{\ell}$ as well as introduce the new field, $\phi^Z$, defined as
\begin{equation}
Z\left(y\right) = y^{-\frac{1}{D - 3}} \phi^{\text{Z}}\left(y\right).
\end{equation}
With this, the equation of motion simplifies to
\begin{equation}
\label{eq:EoMZerilliPhi}
\partial_y\left(\Delta \partial_y \phi^{\text{Z}}\left(y\right)\right) + \partial_y \phi^{\text{Z}}\left(y\right) - \hat{\ell}\left(\hat{\ell} + 1\right) \phi^{\text{Z}}\left(y\right) = 0.
\end{equation}
The resulting ladder operators are\footnote{Note that the method of \autoref{a:HypergeometricIdentities} cannot be used here as the form of the parameters $a$, $b$ and $c$ of the hypergeometric function solution to \autoref{eq:EoMZerilliPhi} is not of the form assumed when deriving \autoref{eq:RaisingOperator} and \autoref{eq:LoweringOperator}. Instead, we find it easiest to bootstrap the raising and lowering operators based on their known algebra with the Hamiltonian, \autoref{eq:LadderOperatorAlgebra}.}
\begin{subequations}
\begin{align}
\label{eq:Spin2ScalarRaising}
D_{\hat{\ell}}^+ &= -\Delta \partial_y - \frac{\left(\hat{\ell} + 1\right) \Delta' + 1}{2},\\
D_{\hat{\ell}}^- &= \Delta \partial_y - \frac{\hat{\ell} \Delta' - 1}{2}.
\label{eq:Spin2ScalarLowering}
\end{align}
\end{subequations}
As before, we can demonstrate that these operators are indeed true raising and lowering operators by computing their intertwining relation and algebra with the Hamiltonian, defined as
\begin{equation}
\label{eq:HamiltonianSpin2Scalar}
H_{\hat{\ell}} \equiv -\Delta\left[\partial_y\left(\Delta \partial_y\right) + \partial_y - \hat{\ell}\left(\hat{\ell} + 1\right)\right].
\end{equation}
In terms of the ladder operators, the Hamiltonian is expressed as
\begin{subequations}
\begin{align}
\label{eq:Spin2ScalarHamiltonian}
H_{\hat{\ell}} &= D_{\hat{\ell} - 1}^+ D_{\hat{\ell}}^- - \frac{\hat{\ell}^{\:2} - 1}{4},\\
&= D_{\hat{\ell} + 1}^- D_{\hat{\ell}}^+ - \frac{\hat{\ell}\left(\hat{\ell} + 2\right)}{4}.
\end{align}
\end{subequations}
The intertwining relation can be computed to be
\begin{equation}
D_{\hat{\ell} + 1}^- D_{\hat{\ell}}^+ - D_{\hat{\ell} - 1}^+ D_{\hat{\ell}}^- = \frac{2\hat{\ell} + 1}{4}.
\end{equation}
The same ladder operator algebra as \autoref{eq:LadderOperatorAlgebra} then holds for these definitions.

Examining \autoref{eq:Spin2ScalarHamiltonian} reveals that the ground level is $\hat{\ell} = 1$\footnote{Since the ladders raise and lower by integer values of $\hat{\ell}$, the branch of solutions that starts at $\hat{\ell}=-1$ overlaps with the positive branch of solutions for physical values of $\ell$. We therefore exclude this second branch of solutions as there cannot be multiple solutions for a given $\ell$.}, at which a ground state can be found:
\begin{equation}
H_{1} \phi^{\text{Z}}_{\hphantom{\text{Z}} 1} = D_{0}^+ D_{ 1}^- \phi^{\text{Z}}_{\hphantom{\text{Z}} 1} = 0.
\end{equation}
As emphasized in the previous subsection, although the ladder connects to all integer values of $\hat{\ell}$, only those multipoles with $\ell \geq 2$ are tidal perturbations. We therefore find the first-order differential equation
\begin{equation}
D_{1}^{-}\phi_{\ 1}^{\text{Z}}=0,
\end{equation}
which can be solved to give
\begin{equation}
\phi^{\text{Z}}_{\hphantom{\text{Z}}1} \propto y.
\end{equation}
As $y \rightarrow 1$, $\phi^{\text{Z}}_{\hphantom{\text{Z}}1}\left(y\right)$ is regular at the horizon. Acting with the raising operator repeatedly to connect to solutions at higher $\hat{\ell}$ does not introduce a decaying tail. Thus, we conclude that the Love numbers vanish for this branch of solutions, i.e., for integer $\hat{\ell}$.

This is in agreement with \cite{Kol:2011}, where they compute the Love numbers for what we call $\phi^{Z}$, and \cite{Hui:2020} who track the definition of the Love number back through the various field redefinitions to compute the Love number for the original metric perturbation master variable $\Psi^{\text{Z}}$. Indeed, the authors of \cite{Hui:2020} show that the two definitions of the spin 2 scalar Love number, the one corresponding to the field $\phi^Z$, which we denote $\lambda_{\phi}$, and the one corresponding to the field $\Psi^{\text{Z}}$, denoted $\lambda_{\Psi}$, are related by an overall constant:
\begin{equation}
\lambda_{\Psi}= -\frac{\left(\ell +D-3\right)\left(\ell +D-2\right)^{2}}{\ell \left(\ell -1\right)^{2}} \lambda_{\phi}.
\end{equation}
Hence, if one Love number vanishes, then so does the other. In fact, the gauge invariance of spin 2 perturbations means that there is no one preferred variable for which to define the black hole's static response: each choice of even parity variable is equally valid. We therefore work with the simpler variable whose equation of motion allows us to straightforwardly write down ladder operators.

\section{Ladders for 5D Myers--Perry}
\label{s:MyersPerry}
In higher dimensions, a whole zoo of black hole solutions to Einstein's equations emerges. Recent work has explored the tidal response of these higher-dimensional black holes (e.g. \cite{Rodriguez:2023, Glazer:2024}) or solutions of general fields on fixed Myers--Perry backgrounds (e.g. \cite{Lunin:2017, Lunin:2025}). In doing so, one finds that there is often no known analytic solution of the equations of motion for perturbations to these black holes for all regions of spacetime. The ladder symmetry analysis developed in this paper may nevertheless be useful for studying consequences of these equations since the ladder operators obey a known algebra with the Hamiltonian and can, in principle, be found without solving the full equations of motion. Our constructive arguments therefore provide a direct test for the vanishing of Love numbers. As a proof of concept, we study scalar perturbations to the 5D Myers--Perry black hole, which is described by the following line element in the analog of Boyer-Lindquist coordinates \cite{Myers:1986}:  
\begin{equation}
\label{eq:MyersPerryMetric}
\begin{aligned}
\mathrm{d}s^{2}= & -\mathrm{d}t^{2}+\frac{\mu}{\Sigma}\left(\mathrm{d}t-a\sin^{2}\theta\mathrm{d}\phi-b\cos^{2}\theta\mathrm{d}\psi\right)^{2}+\frac{r^{2}}{\bar{\Delta}}\mathrm{d}r^{2}+\Sigma\mathrm{d}\theta^{2}\\
 & +\left(r^{2}+a^{2}\right)\sin^{2}\theta\mathrm{d}\phi^{2}+\left(r^{2}+b^{2}\right)\cos^{2}\theta\mathrm{d}\psi^{2},
\end{aligned}
\end{equation}
where, for free parameters $\mu$, $a$, and $b$,
\begin{subequations}
\begin{align}
\Sigma & \equiv r^{2}+a^{2}\cos^{2}\theta+b^{2}\sin^{2}\theta,\\
\bar{\Delta} & \equiv\left(r^{2}+a^{2}\right)\left(r^{2}+b^{2}\right)-\mu r^{2}.
\end{align}
\end{subequations}
This is a natural choice for a first step into the higher-dimensional landscape since it is an exactly solvable model that has a well-defined 4D limit. Furthermore, it is known that, for otherwise arbitrary parameters, there is no choice of $\hat{\ell}$ for which the Love number vanishes \cite{Rodriguez:2023, Glazer:2024}, giving a more robust test of our method than Schwarzschild--Tangherlini. In this section we demonstrate that higher-dimensional ladder structures are more ubiquitous than just spherically symmetric black holes, and that they are able to straightforwardly diagnose the specific tuning of parameters needed to allow for a vanishing Love number. We first derive the ladder operators and then use them in combination with the ground-state argument to demonstrate when the Love numbers vanish. A review of the harmonic decomposition and derivation of the equation of motion, following \cite{Rodriguez:2023}, is done in \autoref{a:MyersPerry}.

The radial part of the Klein--Gordon equation for a scalar perturbation to 5D Myers--Perry, \autoref{eq:EoM5DMyersPerry}, is
\begin{equation}
\label{eq:MyersPerryEom}
\partial_{x}\left[\left(x^{2}-1\right)\partial_{x}R_{\hat{\ell}}\right]+2\left(\frac{\left(\tilde{m}_{L}+\tilde{m}_{R}\right)^{2}}{x-1}-\frac{\left(\tilde{m}_{L}^{2}-\tilde{m}_{R}^{2}\right)^{2}}{x+1}\right)R_{\hat{\ell}} - \hat{\ell}\left(\hat{\ell}+1\right)R_{\hat{\ell}} = 0,
\end{equation}
where $x$ is related to the radial variable by
\begin{equation}
r^2 = \frac{\left(r_+^2 - r_-^2\right) x + r_+^2 + r_-^2}{2}
\end{equation}
for horizons
\begin{equation}
r_{\pm}^2 = \frac{1}{2}\left(\mu - a^2 - b^2 \pm \sqrt{\left(\mu - a^2 - b^2\right)^2 - 4a^2 b^2}\right),
\end{equation}
so that the inner and outer horizons are located at $x = \pm 1$. Additionally, $\tilde{m}_L$ and $\tilde{m}_R$ are related to the magnetic quantum numbers associated to $\phi$ and $\psi$, denoted by $m_\phi$ and $m_\psi$ respectively, as:
\begin{subequations}
\begin{align}
\label{eq:mphisec}
m_\phi &= 2\left(\frac{r_+ - r_-}{a + b} \tilde{m}_R + \frac{r_+ + r_-}{a - b} \tilde{m}_L\right),\\
m_\psi &= 2\left(\frac{r_+ - r_-}{a + b} \tilde{m}_R - \frac{r_+ + r_-}{a - b} \tilde{m}_L\right).
\label{eq:mpsisec}
\end{align}
\end{subequations}

The ladder operators associated to this equation of motion are
\begin{subequations}
\begin{align}
D_{\hat{\ell}}^+ &= -\Delta \partial_x - \frac{\left(\hat{\ell} + 1\right)\Delta'}{2} - \frac{4\tilde{m}_L \tilde{m}_R}{\hat{\ell} + 1},\\
\label{eq:MyersPerryLowering}
D_{\hat{\ell}}^- &= \Delta \partial_x - \frac{\hat{\ell} \Delta'}{2} + \frac{4\tilde{m}_L \tilde{m}_R}{\hat{\ell}},
\end{align}
\end{subequations}
where, in analogy to the Schwarzschild--Tangherlini cases, we have defined
\begin{equation}
\Delta\equiv\left(x^{2}-1\right),
\end{equation}
not to be confused with $\bar{\Delta}$ appearing in the metric, \autoref{eq:MyersPerryMetric}. These ladder operators have the intertwining relation
\begin{equation}
D_{\hat{\ell}+1}^{-}D_{\hat{\ell}}^{+}-D_{\hat{\ell}-1}^{+}D_{\hat{\ell}}^{-}=2\hat{\ell}+1-\frac{16\left(2\hat{\ell}+1\right)\tilde{m}_{L}^{2}\tilde{m}_{R}^{2}}{\hat{\ell}^{\:2}\left(\hat{\ell}+1\right)^{2}}.
\end{equation}
Defining the Hamiltonian
\begin{equation}
H_{\hat{\ell}} \equiv -\Delta \left[\partial_{x}\left(\Delta\partial_{x}\right) + 2\left(\frac{\left(\tilde{m}_{L}+\tilde{m}_{R}\right)^{2}}{x-1} - \frac{\left(\tilde{m}_{L} - \tilde{m}_{R}\right)^{2}}{x+1}\right) - \hat{\ell}\left(\hat{\ell}+1\right)\right],
\end{equation}
one finds
\begin{subequations}
\begin{align}
\label{eq:HMyersPerryDminus}
H_{\hat{\ell}} & =D_{\hat{\ell}-1}^{+}D_{\hat{\ell}}^{-}-\frac{\left(\hat{\ell}^{\:2}+4\tilde{m}_{L}^{2}\right)\left(\hat{\ell}^{\:2}+4\tilde{m}_{R}^{2}\right)}{\hat{\ell}^{\:2}}\\
\label{eq:HMyersPerryDplus}
& =D_{\hat{\ell}+1}^{-}D_{\hat{\ell}}^{+}-\frac{\left(\left(\hat{\ell}+1\right)^{2}+4\tilde{m}_{L}^{2}\right)\left(\left(\hat{\ell}+1\right)^{2}+4\tilde{m}_{R}^{2}\right)}{\left(\hat{\ell}+1\right)^{2}}
\end{align}
\end{subequations}
along with the usual ladder operator algebra \autoref{eq:LadderOperatorAlgebra}.

Given the Schwarzschild--Tangherlini analysis, if the Love number vanishes for 5D Myers--Perry then the one expects that the $D_{\hat{\ell}-1}^{+}D_{\hat{\ell}}^{-}$ form of the equation of motion \autoref{eq:HMyersPerryDminus} will lead to the regular solution ground state. On the other hand, the expectation is that the $D_{\hat{\ell}+1}^{-}D_{\hat{\ell}}^{+}$ form of the equation of motion \autoref{eq:HMyersPerryDplus} will lead to the irregular solution ground state (see \autoref{a:GroundStateForDecayingSolutions}); otherwise the Schwarzschild limit will not be smooth. However, we encounter a problem that was not present in the Schwarzschild--Tangherlini cases: the Hamiltonian in \autoref{eq:HMyersPerryDminus} and the lowering operator in \autoref{eq:MyersPerryLowering} are undefined for $\hat{\ell} = 0$. It is clear, however, from the equation of motion, \autoref{eq:MyersPerryEom}, that there is a valid $\hat{\ell} = 0$ solution. Indeed, we expect that, for spin 0 perturbations, all multipoles are tidal perturbations. Our ground-state argument therefore breaks down for spin 0 perturbations to 5D Myers--Perry.

The source of the problem is the constant term $16\tilde{m}_{L}^{2}\tilde{m}_{R}^{2}/\hat{\ell}^{\:2}$ in \autoref{eq:HMyersPerryDminus}. This term, which is the roadblock to our ground state analysis, only vanishes for $\tilde{m}_{L}=0$ or $\tilde{m}_{R}=0$. Following \autoref{eq:mphisec}--\autoref{eq:mpsisec}, we see that the two conditions under which this occurs are either $\abs{a} = \abs{b}$ or $\abs{m_{\phi}} = \abs{m_{\psi}}$.

By inspection, we recover Schwarzschild--Tangherlini for $\tilde{m}_{L} = \tilde{m}_{R}=0$. On the other hand, with $\tilde{m}_{L} = 0$ and $\tilde{m}_{R} \neq 0$ (and vice versa) we find that the  Hamiltonian takes the form
\begin{equation}
\tilde{H}_{\hat{\ell}} = -\Delta\left[\partial_x\left(\Delta \partial_x\right) + \frac{4\tilde{m}_R^2}{\Delta} - \ell\left(\ell + 1\right)\right],
\end{equation}
where we put a tilde on $H$ and later $D_{\hat{\ell}}^{\pm}$ to indicate that this is valid when $\tilde{m}_L = 0$. Notice that, in this limit, the Hamiltonian has a similar form to the 4D Kerr Hamiltonian for scalar perturbations \cite{Hui:2021}. Using this as inspiration, we employ the following field re-definition:
\begin{equation}
R_{\hat{\ell}} = \left(\frac{x - 1}{x}\right)^{2i\tilde{m}_R} \psi_{\hat{\ell}}, 
\end{equation}
which gives us a new equation of motion with associated Hamiltonian 
\begin{equation}
\tilde{H}_{\hat{\ell}} = -\Delta\left[\partial_x\left(\Delta \partial_x\right) + 4 i \tilde{m}_R - \ell\left(\ell + 1\right)\right].
\end{equation}
The new field redefinition is not merely a trick. Instead, the field $\psi$ is the proper variable on which to impose the boundary condition of regularity at the horizon. One can check this since for $\psi \sim$ constant, $T_{\mu\nu} u^\mu u^\nu$ is finite at the horizon, where $T_{\mu\nu}$ is the stress-energy tensor for the scalar and $u^\mu$ is the four-velocity of a freely-falling observer. On the other hand, had we stuck with $R_{\hat{\ell}}$, the solution for which the energy density measured by a freely-falling observer is finite would correspond to a divergent $R_{\hat{\ell}}$. With that said, in this form, the ladder operators become
\begin{subequations}
\begin{align}
\tilde{D}_{\hat{\ell}}^{+} &= -\Delta \partial_x - \frac{\hat{\ell} + 1}{2} \Delta' - 2i \tilde{m}_R,\\
\tilde{D}_{\hat{\ell}}^- &= \Delta \partial_x - \frac{\hat{\ell}}{2} \Delta' + 2 i \tilde{m}_R.
\end{align}
\end{subequations}
The Hamiltonian can once again be expressed in terms of these new ladder operators 
\begin{subequations}
\begin{align}
\tilde{H}_{\hat{\ell}} &= \tilde{D}_{\hat{\ell} - 1}^+ \tilde{D}_{\hat{\ell}}^- - \left(\frac{\hat{\ell}^2}{4} + 4 \tilde{m}_R^2\right), \\
&= \tilde{D}_{\hat{\ell} + 1}^- \tilde{D}_{\hat{\ell}}^+ - \left(\frac{\left(\hat{\ell} + 1\right)^2}{4} + 4\tilde{m}_R^2\right).
\end{align}
\end{subequations}

Cast in this form, we are able to make a constructive argument once again for the vanishing of Love numbers. Observe that the Hamiltonian factorizes as
\begin{equation}
\label{eq:FactorizedMP}
\tilde{H}_{\hat{\ell}} = \left(\tilde{D}^+_{\hat{\ell} - 1} - 2i\tilde{m}_R\right)\left(\tilde{D}^-_{\hat{\ell}} - 2i \tilde{m}_R\right) - \frac{\hat{\ell}^{\:2}}{4} - 2i\tilde{m}_R \hat{\ell} \: \Delta'. 
\end{equation}
It is clear now that the ground level is $\hat{\ell} = 0$, where the Hamiltonian reduces to
\begin{equation}
\tilde{H}_0 \psi_0 = \left(\tilde{D}^+_{- 1} - 2i\tilde{m}_R\right)\left(\tilde{D}^-_{0} - 2i \tilde{m}_R\right)\psi_0 = 0.
\end{equation}
From this, we obtain the first order differential equation
\begin{equation}
\left(\tilde{D}^-_{\hat{\ell}} - 2i\tilde{m}_R\right) \psi_0 = \Delta \partial_x \psi_0 = 0.
\end{equation}
The solution is $\psi_0 =$ constant, and we run the same arguments as before to conclude that acting repeatedly with the raising operator $D_{\hat{\ell}}^{+}$ will produce only modes with purely growing behavior as $r \rightarrow \infty$. As a result, the Love number vanishes for all integer $\hat{\ell}$, again remembering the additional condition $\tilde{m}_{L} = 0$ and $\tilde{m}_{R} \neq 0$ (and vice versa).

Note that, although the form of the ladders and Hamiltonian presented above are reminiscent of those for a 4D Kerr black hole presented in \cite{Hui:2021}, the argument for why the tidal Love numbers vanish is different. Here, instead of using horizontal ladder symmetries constructed out of the Wronskian (see \autoref{a:Horizontal}), we notice that the factorization in \autoref{eq:FactorizedMP} allows for the use of the constructive ground state procedure. 

We have therefore shown that the ground-state argument breaks down for scalar perturbations to 5D Myers--Perry, suggesting that the Love numbers do not generically vanish. However, for $\tilde{m}_{L}=0$ or $\tilde{m}_{R}=0$, corresponding to either $\abs{a}=\abs{b}$ or $\abs{m_{\phi}}=\abs{m_{\psi}}$, the problematic term in \autoref{eq:HMyersPerryDminus} is zero and we can find a ground state. From this ground level, we climb the ladder and conclude that the Love numbers vanish for all integer values of $\hat{\ell}$. This is in agreement with the results of \cite{Charalambous:2023, Rodriguez:2023}, who showed that the Love numbers for scalar perturbations to 5D Myers--Perry are non-vanishing unless both $\hat{\ell}=\text{integer}$ and either $\abs{a}=\abs{b}$ or $\abs{m_{\phi}}=\abs{m_{\psi}}$. This demonstrates that the ladder operator/ground state argument is useful not only for finding those multipoles at which the Love number does vanish, but also as a test of the question of whether the Love number can vanish.

\section{Discussion}
\label{s:Discussion}

In this paper, we demonstrate that the ladder argument of 
\cite{Hui:2021, Berens:2022} (used in conjunction with the existence of a ground state) generalizes to higher dimensions, providing a symmetry rationale for the vanishing of black hole static Love numbers for specific values of the angular momentum quantum number\footnote{The connection between the ladder operators, and the ladder symmetry that encodes them, is explained around \autoref{verticalSymm}.}. For general dimensions, the relevant quantity is an effective level number $\hat \ell \equiv \ell / (D-3)$ where $D$ is the number of spacetime dimensions. For perturbations to Schwarzschild--Tangherlini we show that, {\it when} there exists a ground level (such that the equation of motion can be reduced to a first order equation for the ground state), the Love number for any $\hat \ell$ that is connected to the ground level by a positive integer vanishes, matching what has been found in the literature. This symmetry reasoning is constructive: in the 5D scalar Myers--Perry case for which the Love number is generically non-vanishing, no appropriate ground state can be found unless one tunes the magnetic quantum numbers (related to the black hole spins) to the precise values that have been shown to be needed for vanishing Love numbers. A summary of the main results is given in \autoref{tab:Results}.

\begin{table}[t]
\centering
 \begin{tabular}{||c | c | c | c | c ||} 
 \hline
  & Perturbation & Ground Level & Vanishing Love Numbers \\ [0.5ex] 
 \hline\hline
 \multirow{5}{7em}{Schwarzschild--Tangherlini} & Spin 0 & $\hat{\ell} = 0$ & $\hat{\ell} \in \mathbb{Z}_{\geq 0}$ \\ 
  & Spin 1 Odd &  $\hat{\ell} = \pm \frac{1}{D - 3}$ & $\hat{\ell} = n \pm \frac{1}{D - 3},\:\: n\in \mathbb{Z}_{\geq 0}$ \\
  & Spin 1 Even &  $\hat{\ell} = 1$ & $\hat{\ell} \in \mathbb{Z}_{\geq 1}$ \\
  & Spin 2 Vector & $\hat{\ell} = \pm \frac{D - 2}{D - 3}$ & $\hat{\ell} = n \pm \frac{D - 2}{D - 3}, \: \: n \in \mathbb{Z}_{\geq 0}$ \\
  & Spin 2 Scalar &  $\hat{\ell} = 1$ & $\hat{\ell} \in \mathbb{Z}_{\geq 2}$ \\ 
  \hline
 5D Myers--Perry & Spin 0 & $\hat{\ell} = 0$ and either $\tilde{m}_L = 0$ or $\tilde{m}_R = 0$ & $\hat{\ell} \in \mathbb{Z}_{\geq 0}$ \\ [1ex] 
 \hline
 \end{tabular}
 \caption{Summary of results. The ground level determines the location of the ground state while the last column says for which values of $\hat{\ell}$ the Love numbers vanish. As can be seen, the Love numbers are zero for multipoles connected to the ground level by an integer, because they are connected through the ladder operators. We emphasize that the only physical values of $\ell = \left(D - 3\right)\hat{\ell}$ are $\ell\geq0$, and the only values corresponding to tidal perturbations are $\ell \geq s$, where $s = \left\{0, 1, 2\right\}$ is the spin of the perturbation.}
\label{tab:Results}
\end{table}

The fact that a ladder structure exists in higher dimensions raises the question of whether it has a geometric origin. This is known to be the case for spin 0, 1, 2 perturbations around a 4D Schwarzschild black hole \cite{Berens:2025okm}. There, the solutions organized by angular momentum quantum numbers form a principal series representation of SO(3,1), which arises from 3 Killing vectors and 3 conformal Killing vectors which obey a special condition known as the melodic condition \cite{Berens:2022}. It can be shown that this special condition is not satisfied in higher dimensions in general. Nonetheless, it is tempting to think that the ladder structure in higher dimensions arises from the representation theory of some symmetry group. 
Exactly what this could be is a subject of current investigation. Uncovering this could also pave the way to identify the correct symmetries at the level of the worldine EFT, similar to what has been done in 4D \cite{Berens:2025okm}. 

Studying higher-dimensional ladders could shed light on what happens in other situations where the Love numbers do not generically vanish, such as for other compact objects like neutron stars, dynamical perturbations or, as has been recently shown, for fermionic perturbations \cite{Chakraborty:2025}. Another avenue for future work would be to utilize symmetries such as the ones found here to explore conjectured relationships, including the equivalence of Love numbers in $(D - 1)$-dimensional Myers--Perry and $D$-dimensional black rings and black strings, as was noticed in \cite{Rodriguez:2023}.

\paragraph{Acknowledgments.} We would like to thank Austin Joyce, Riccardo Penco, Massimiliano M. Riva, and Luca Santoni for many useful discussions throughout the course of this project. The work of LH, DM and JS is supported by the US Department of Energy grant DE-SC011941. The work of RB is supported by the NSF CAREER award PHY-2340457 and the Simons Foundation award SFI-MPS-BH-00012593.

\newpage

\appendix

\section{Horizontal Ladder Symmetry}
\label{a:Horizontal}

An alternative argument for the vanishing of the Love numbers can be phrased in terms of a different symmetry of the Hamiltonian. The key ingredient is the Wronskian, which determines whether any two solutions to a differential equation are linearly independent. Its utility stems from the fact that it is conserved on-shell. The fact that the Wronskian, and its powers, are conserved was used in \cite{BenAchour:2022} to precisely describe the symmetry structure associated to linear perturbations around a Schwarzschild black hole. 

For our purposes, we will fix one of the arguments of the Wronskian to be the solution of the equation of motion that is regular at the horizon, 
\begin{equation}
Q_{\hat{\ell}_{g}} = W\left[\phi_{\hat{\ell}_{g}}^{\text{reg}}, \cdot\right],
\end{equation}
where $\phi_{\hat{\ell}_{g}}$ denotes the solution at the ground level. Similar operators for any level connected to $\hat{\ell}_{g}$ by an integer can be obtained using the ladder operators. For example, if the ground level is $\hat{\ell}_{g}=0$, then
\begin{equation}
Q_{\hat{\ell}} = D_{\hat{\ell} - 1}^+ \dots D_0^+ Q_0 D_1^- \dots D_{\hat{\ell}}^-.
\end{equation}
The conserved quantity $P_0 \equiv Q_0 \phi_0$ is independent of $r$. For a single solution, $P_{0}$ must therefore have the same value at $r \rightarrow \infty$ as it does at $r \rightarrow r_s$. This is useful for the vanishing of Love numbers problem as it connects the asymptotics of the two branches of solutions. By construction, $P_0$ is zero when a solution is regular at the horizon and non-zero when it is irregular at the horizon. If the operator $Q_{\hat{\ell}}$, obtained with the help of the ladder operators, is zero when applied to the mode that is growing at $r\rightarrow\infty$, then the growing mode can only asymptote to be regular at the horizon. Similarly, if $Q_{\hat{\ell}}$ is non-zero when applied to the mode that decays at $r\rightarrow\infty$, then the decaying solution must become irregular near to the horizon. In other words, while the regular mode at the horizon can in principle asymptote to a linear combination of a growing and decaying modes at infinity, the Wronskian shows whether or not it only asymptotes to one of these. The symmetry stated here was introduced in \cite{Hui:2021} and dubbed a \textit{horizontal ladder symmetry} since it operates at each level of the ladder independently and because $\delta \phi_{\hat{\ell}} = Q_{\hat{\ell}} \phi_{\hat{\ell}}$ is a symmetry of the action for the decomposed $\phi_{\hat{\ell}}$ \cite{Hui:2021,Berens:2022}. 

\subsection{Spin 0}
We now apply this analysis to the case of spin 0 scalar perturbations to Schwarzschild--Tangherlini discussed in \autoref{s:Spin0}. For \autoref{eq:Spin0EOM}, the Wronskian is
\begin{equation}
W\left[\phi_{\hat{\ell}}^a\left(y\right), \phi_{\hat{\ell}}^b\left(y\right)\right] = \phi_{\hat{\ell}}^a \Delta \partial_y \phi_{\hat{\ell}}^b - \phi_{\hat{\ell}}^b \Delta \partial_y \phi_{\hat{\ell}}^a. 
\end{equation}
The solution at $\hat{\ell} = 0$ that is regular at the horizon is simply a constant. It follows that
\begin{equation}
Q_0 = \Delta \partial_y. 
\end{equation}
Since the Hamiltonian at $\hat{\ell} = 0$ is $H_0 = - Q_0^2$, 
\begin{equation}
\comm{Q_0}{H_0} = 0.
\end{equation}
Climbing the ladder, one can then find that
\begin{equation}
Q_{\hat{\ell}} \equiv D_{\hat{\ell} - 1}^+ Q_{\hat{\ell} - 1} D_{\hat{\ell}}^- \Rightarrow \comm{Q_{\hat{\ell}}}{H_{\hat{\ell}}} = 0.
\end{equation}
The associated conserved quantity at the ground level $\hat{\ell} = 0$ is
\begin{equation}
P_0 \equiv Q_0 \phi_0 \Rightarrow \partial_y P_0 = 0,
\end{equation}
from which the level $\hat{\ell}$ conserved quantity,  $\partial_y P_{\hat{\ell}} = 0$, is given through the use of ladder operators:
\begin{equation}
P_{\hat{\ell}} \equiv \Delta \partial_y \left(D_1^- D_2^- \dots D_{\hat{\ell}}^- \phi_{\hat{\ell}}\right).
\end{equation}

To be clear, we can only input solutions into the Wronskian that are of the same $\hat{\ell}$ level. In order to define conserved quantities for higher $\hat{\ell}$, we apply a series of lowering operators to yield the $\hat{\ell} = 0$ solution and apply what we found to be conserved at that level.

Note that the conserved quantity defined here, $P_{\hat{\ell}}$, is not exactly the Noether charge associated to the symmetry generated by $Q_{\hat{\ell}}$. The Noether charge is actually $P_{\hat{\ell}}$ squared. Indeed, as described above, $P_{\hat{\ell}}$ is constructed such that its value associated to the solution regular at the horizon is zero while its value associated to the solution irregular at the horizon is non-zero. Far away from the black hole, one can write the solutions to the equations of motion as a linear combination of two functions, one that is growing, $\phi_{\hat{\ell}} \sim y^{\hat{\ell}}$, and one that is decaying, $\phi_{\hat{\ell}} \sim y^{-\hat{\ell} - 1}$. Plugging these solutions into $P_{\hat{\ell}}$, we find that the charge evaluated on the growing mode vanishes, but when evaluated on the decaying mode it does not. As such, the decaying mode must include a piece that is irregular at the horizon. This additionally tells us that the growing mode must approach a regular solution at the horizon. Therefore, by imposing regularity at the horizon, we must discard the decaying tail at infinity, while retaining the growing mode. This is equivalent to saying the Love numbers vanish. 

\subsection{Spin 1}
We now demonstrate that the horizontal ladder symmetry analysis is useful for all spins, starting with spin 1 perturbations to Schwarzschild--Tangherlini. 

\subsubsection{Vector Perturbations}
As was the case for spin 0, one can express the vanishing of the Love numbers for the spin 1 vector master variable of \autoref{s:OddParityDegreeOfFreedom} in terms of a horizontal ladder symmetry of the corresponding Hamiltonian \autoref{eq:HamiltonianSpin1Odd}. In the spin 0 case, it was helpful to start at level $\hat{\ell} = 0$. In this case, as shown in \autoref{eq:Spin1OddReducedEoM}, the simplest level to start with is the ground level $\hat{\ell} = \pm \mu$. There, the Hamiltonian becomes
\begin{equation}
H_{\pm \mu}\phi^V_{\hphantom{V}\pm \mu} = -\left(y - 1\right) y^{2 \mp \mu} \partial_y \left(\left(y - 1\right) y^{1 \pm 2\mu} \partial_y \left(y^{-\left(\pm \mu + 1\right)} \phi^V_{\hphantom{V}\pm \mu}\right)\right).
\end{equation}
While we can immediately read off a conserved quantity, we do not know a priori that this is the most useful conserved quantity for our horizontal ladder symmetry arguments. We instead want to use the Wronskian since we can, by construction, write a conserved quantity that is zero for the mode regular at the horizon and non-zero for the mode irregular at the horizon, which may not be the same as the conserved quantity found by inspection. The Wronskian for two solutions at this level $\hat{\ell} = \pm \mu$, which we denote by superscript $a$ and $b$, is given by
\begin{equation}
W\left[\phi^{V, a}_{\pm \mu}, \phi^{V, b}_{\pm \mu}\right] = \frac{y - 1}{y} \left(\phi^{V, a}_{\pm \mu} \partial_y \phi^{V, b}_{\pm \mu} - \phi^{V, b}_{\pm \mu} \partial_y \phi^{V, a}_{\pm \mu}\right).
\end{equation}
Plugging in the regular solution \autoref{eq:Spin1VReg}, we find the following conserved quantity
\begin{equation}
P_{\pm \mu} \equiv W\left[\phi^{V, \text{reg}}_{\pm \mu}, \phi^V_{\pm\mu}\right] = \left(y - 1\right)y^{1 \pm 2 \mu} \partial_y \left(y^{-\left(\pm \mu + 1\right)}\phi^V_{\pm \mu}\right).
\end{equation}
We can define the symmetry transformation associated to this conserved quantity as
\begin{equation}
Q_{\pm \mu} \phi_{\pm \mu}^V \equiv y^{\pm \mu + 1} \partial_y\left(y^{-\left(\pm \mu + 1\right)} \phi_{\pm}^V\right).
\end{equation}
The Hamiltonian rescaled by $y^{\pm 2\mu}$ commutes with $Q_{\pm \mu}$ since $H_{\pm \mu} = y^{\mp 2\mu} Q_{\pm \mu}^2$, 
\begin{equation}
\comm{Q_{\pm \mu}}{y^{\pm 2\mu} H_{\pm \mu}} = 0.
\end{equation}
Since rescaling the Hamiltonian does not change the solutions to the equations of motion, this demonstrates that $Q_{\pm \mu}$ is a symmetry of those equations of motion. At this point, one can write down conserved quantities at any level $\hat{\ell} = n \pm \mu$ for $n \in \mathbb{Z}_{\geq 0}$ by using lowering operators to connect back to the ground level solution at $\hat{\ell} = \pm \mu$, 
\begin{equation}
P_{\hat{\ell}} = \left(y - 1\right) y^{1 \pm 2\mu} \partial_y\left(y^{-\left(\pm \mu + 1\right)} D_1^- D_2^- \dots D_{\hat{\ell}}^- \phi_{\hat{\ell}}\right).
\end{equation}
The argument now runs as before. The conserved quantity $P_{\hat{\ell}}$ is zero at the horizon. Then, asymptotically far away, at level $\hat{\ell}$, \autoref{eq:hatlSpin1V}, the solutions can be taken to either be growing or decaying at infinity. Plugging in the growing mode in to $P_{\hat{\ell}}$ also yields $0$, hence the growing mode is constructed out of the mode that is regular at the horizon. Meanwhile, the decaying mode is nonzero, hence it must connect to the mode that is irregular at the horizon. This is equivalent to saying the Love numbers vanish for those $\hat{\ell}$ in \autoref{eq:hatlSpin1V}, as before. 

\subsubsection{Scalar Perturbations}
For spin 1 scalar perturbations to Schwarzschild--Tangherlini, \autoref{s:EvenParityDegreeOfFreedom}, an equivalent story in terms of horizontal symmetries of \autoref{eq:HamiltonianSpin1Even} starts at the spin 1 scalar ground level $\hat{\ell} = 1$. The conserved charge at this level is 
\begin{equation}
P_1 \equiv y^3\left(y - 1\right)\partial_y\left(y^{-2}\phi_1\left(y\right)\right),
\end{equation}
with the associated symmetry generator,
\begin{equation}
Q_1 \equiv y^5 \left(y - 1\right)\partial_y\left(y^{-1} \phi^S_1\right).
\end{equation}
This commutes with a rescaling of the Hamiltonian \autoref{eq:HamiltonianSpin1Even},
\begin{equation}
\comm{Q_1}{y^4 H_1} = 0.
\end{equation}
Conserved charges at all level $\hat{\ell}$ connected to $\hat{\ell} = 1$ by an integer are given by
\begin{equation}
P_{\hat{\ell}} = y^3\left(y - 1\right)\partial_y\left(y^{-2} D_1^- D_2^- \dots D_{\hat{\ell}}^- \phi_{\hat{\ell}}\left(y\right)\right).
\end{equation}
Once again, the conserved charge is constructed from the Wronskian. Therefore, the conserved charge, when applied to the mode that is regular at the horizon, is zero and is non-zero for the mode that is irregular at the horizon. Then, one can use the conserved charge to study what happens asymptotically far away from the black hole, where the solution can be cast in terms of a mode that is growing and a mode that is decaying. The growing mode returns zero for the conserved charge, while the decaying mode yields a non-zero result. Consequently, by charge conservation, we find that the Love numbers vanish for all integer $\hat{\ell}$. 

\subsection{Spin 2}
For spin 2 perturbations to Schwarzschild--Tangherlini the horizontal symmetry analysis also holds. As explained in \autoref{s:TensorPerturbations}, the tensor perturbations reduce to the same equation of motion of that of a spin 0 scalar perturbation, so the arguments carry over immediately from there (with the caveat that spin 2 tidal perturbations begin at $\ell=2$). For vector perturbations, as remarked in \autoref{s:VectorPerturbations}, the spin 2 vector case maps to the spin 1 vector case with the identification $\mu\leftrightarrow\nu$, where $\mu$ and $\nu$ are defined in \autoref{eq:muDefinition} and \autoref{eq:nuDefinition} respectively. From the horizontal ladder symmetry perspective, the symmetry generator can therefore be found by analogy to the vector case in spin 1 as well. In other words, the conserved charge is defined from the Wronskian where one argument is fixed to be the mode regular at the horizon, 
\begin{equation}
P_{\pm \nu} \equiv W\left[\phi_{\pm \nu}^{\text{RW, reg}}, \phi_{\pm \nu}^{\text{RW}}\right] = \left(y - 1\right) y^{1 \pm 2 \nu} \partial_y\left(y^{-\left(\pm \nu + 1\right)} \phi_{\pm \nu}^{\text{RW}}\right).
\end{equation}
The associated symmetry generator is
\begin{equation}
Q_{\pm \nu} \phi^{\text{RW}}_{\pm \nu} = y^{\pm \nu + 1} \partial_y \left(y^{-\left(\pm \nu + 1\right)} \phi_{\pm \nu}^{\text{RW}}\right).
\end{equation}
As before, this commutes with the re-scaled Hamiltonian. At any level $\hat{\ell}$ connected by an integer to $\pm \nu$ is then constructed with the help of the lowering operators
\begin{equation}
P_{\hat{\ell}} = \left(y - 1\right) y^{1 \pm 2 \nu} \partial_y\left(y^{-\left(\pm \nu + 1\right)} D_1^- D_2^- \dots D_{\hat{\ell}}^- \phi_{\hat{\ell}}^{\text{RW}}\right).
\end{equation}
With the charges appropriately defined, the argument runs as before. Since the growing mode at infinity yields zero for the conserved charge, it is matched to the regular mode at the horizon. The decaying mode is non-zero and so it must connect to the irregular mode at the horizon, hence the Love number vanishes for $\hat{\ell} = n \pm \nu$, for $n \in \mathbb{Z}_{\geq 2}$. 

Similarly, the conserved quantity for the ground state Zerilli variable is 
\begin{equation}
P_1 = \Delta^2 \partial_y\left(\frac{\phi_1^{\text{Z}}}{y}\right),
\end{equation}
with the level $\hat{\ell}$ conserved quantity constructed using the ladders as before. The associated symmetry generator at the ground state level is 
\begin{equation}
Q_1 \phi^{\text{Z}}_1 = y\Delta^2 \partial_y\left(\frac{\phi^{\text{Z}}_1}{y}\right).
\end{equation}
As such, the Hamiltonian at this level is $H_1 = -\left(y/\Delta^2\right)Q_1^2$. The level $\hat{\ell}$ solutions can also be constructed using the Zerilli ladders, \autoref{eq:Spin2ScalarRaising} and \autoref{eq:Spin2ScalarLowering}. 

\section{Ground State for Decaying Solutions}
\label{a:GroundStateForDecayingSolutions}

In \autoref{ss:TheVanishingofLoveNumbers}, we ran the ground state argument for spin 0 perturbations to Schwarzschild--Tangherlini. We focused on the $D_{\hat{\ell}-1}^{+}D_{\hat{\ell}}^{-}$ form of the equation of motion in \autoref{eq:HwithDminus}, finding a first-order ODE for $\hat{\ell}=0$ that led to the regular branch of solutions. A natural question is what if instead we had considered the $D_{\hat{\ell}+1}^{-}D_{\hat{\ell}}^{+}$ form of the equation of motion in \autoref{eq:HwithDplus}? Our expectation is that this should produce the branch of solutions that is irregular at the horizon and purely decaying at infinity. We now demonstrate that this is indeed the case in this appendix.

The Hamiltonian for spin 0 perturbations, \autoref{eq:HwithDplus}, can be factorized as:
\begin{equation}
\begin{aligned}
\label{eq:IrregularFactorization}
H_{\hat{\ell}} &= \left(D_{\hat{\ell} + 1}^- - \frac{\Delta'}{2} + \frac{1}{\ln\left(y^2/\Delta\right)}\right)\left(D_{\hat{\ell}}^+ - \frac{\Delta'}{2} + \frac{1}{\ln\left(y^2/\Delta\right)}\right) \\
&\hspace{0.5cm} + \hat{\ell}\left(2\Delta - \frac{\hat{\ell}}{4} - \frac{\Delta'}{\ln\left(y^2/\Delta\right)}\right).
\end{aligned}
\end{equation}
This factorized form is convenient because, for $\hat{\ell}=0$, the equation reduces to the following first-order differential equation of $\phi_0$:
\begin{equation}
\label{eq:ScalarGroundStatePlus}
\left(D_0^+ - \frac{\Delta'}{2} + \frac{1}{\ln\left(y^2/\Delta\right)}\right)\phi_0 = -\left(\Delta \partial_y + \frac{1}{\ln\left(y^2/\Delta\right)}\right)\phi_0 = 0.
\end{equation}
This is a different first-order ODE than \autoref{eq:ScalarGroundStateMinus} and thus constitutes a different ground state solution. The solution to this equation is
\begin{equation}
\phi\left(y\right) \propto \log\left(\frac{y^{2}}{\Delta}\right).
\end{equation}
Since $\Delta\equiv y\left(y-1\right)$, this is irregular at the horizon ($y=1$). Acting repeatedly with the raising operator $D_{\hat{\ell}}^{+}$ and expanding asymptotically reveals that this solution leads to purely decaying behavior at $y\rightarrow\infty$. As is the case in all of our ground state arguments, climbing the ladder with the raising and lowering operators cannot mix between branches of solutions to different first-order differential equations. Therefore, this alternative ground state analysis provides a complementary perspective that the decaying response completely decouples from the growing source. The original equation of motion \autoref{eq:Spin0EOM} can therefore be reduced to two first-order differential equations, \autoref{eq:ScalarGroundStateMinus} and \autoref{eq:ScalarGroundStatePlus}, conveniently written in terms of ladder operators, whose solutions give the two separated branches of regular/growing and irregular/decaying behavior. 

Although the analysis done here was for the case of a spin 0 perturbation, similar decaying-only branches exist for spin 1 and spin 2 and can be constructively found from an irregular ground state.

\section{Harmonic Decomposition of Spin 1 and Spin 2 Perturbations}
\label{a:HarmonicDecompositions}

For the convenience of the reader, we reproduce the derivations of the equations of motion for the master variables corresponding to static spin 1 and spin 2 perturbations to Schwarzschild--Tangherlini done by \cite{Hui:2020}.

\subsection{Spin 1 Harmonic Decomposition}

The physical degrees of freedom for a massless vector field in the Schwarzschild--Tangherlini background can be isolated through a harmonic decomposition. The action is the usual Maxwell action
\begin{equation}
S = -\frac{1}{4} \int F_{\mu\nu} F^{\mu\nu} \sqrt{-g} \dd[D]{x},
\end{equation}
where $F_{\mu\nu}\equiv\partial_{\mu}A_{\nu}-\partial_{\nu}A_{\mu}$ is the field strength tensor in terms of the spin 1 field $A_{\mu}$. The decomposition
\begin{equation}
A_\mu = \begin{pmatrix}
a_0 \\ a_r \\ \partial_i a^{(L)} + a_i^{(T)}
\end{pmatrix}
\end{equation}
splits up the gauge field $A_{\mu}$ into its irreducible representations under the rotation group $SO\left(D-1\right)$ such that $a_{0}$, $a_{r}$ and the longitudinal component $a^{(L)}$ transform as scalars while the transverse component $a_{i}^{(T)}$ transforms as a divergence-free vector ($\nabla^{i}a_{i}^{(T)}=0$), where the subscript $i$ runs over the angular coordinates. In the static limit, we further decompose into hyperspherical harmonics:
\begin{subequations}
\begin{align}
a_0\left(x\right) &= \sum_{\ell, M} a_0\left(r\right) Y_\ell^{M}\left(\theta\right),\\
a_r\left(x\right) &= \sum_{\ell, M} a_r\left(r\right) Y_\ell^{M}\left(\theta\right),\\
a^{(L)}\left(x\right) &= \sum_{\ell, M} a^{(L)}\left(r\right) Y_\ell^{M}\left(\theta\right),\\
a_i^{(T)}\left(x\right) &= \sum_{\ell, M} a^{(T)}\left(r\right) {Y_{i\hphantom{(T)}\ell}^{(T)M}}\left(\theta\right)
\end{align}
\end{subequations}
where ${Y_{i\hphantom{(T)}\ell}^{(T)M}}$ is a divergence-free spin 1 spherical harmonic, $\nabla^{i}{Y_{i\hphantom{(T)}\ell}^{(T)M}}=0$. Using the equations of motion to integrate out the unphysical degrees of freedom, we find the master variable
\begin{equation}
\Psi^{\text{V}}\equiv r^{\frac{D-4}{2}}a^{(T)}
\end{equation}
for the vector sector, while
\begin{equation}
\Psi^{\text{S}}\equiv\frac{r^{D/2}}{\sqrt{\ell\left(\ell+D-3\right)}}a_{0}^{\prime}
\end{equation}
is the scalar master variable, where gauge freedom has been used to set $a^{(L)}=0$. In the static limit, these have the equations of motion
\begin{equation}
\label{eq:Spin1EvenEoM}
f\frac{\mathrm{d}^{2}\Psi^{\text{S}}}{\mathrm{d}r^{2}}+f^{\prime}\frac{\mathrm{d}\Psi^{\text{S}}}{\mathrm{d}r}-\left(\frac{L\left(L+D-3\right)}{r^{2}}+\frac{\left(D-4\right)\left[\left(D-2\right)f-2rf^{\prime}\right]}{4r^{2}}\right)\Psi^{\text{S}}=0
\end{equation}
for the scalar mode and 
\begin{equation}
\label{eq:Spin1OddEoM}
f\frac{\mathrm{d}^{2}\Psi^{\text{V}}}{\mathrm{d}r^{2}}+f^{\prime}\frac{\mathrm{d}\Psi^{\text{V}}}{\mathrm{d}r}-\left(\frac{\left(L+1\right)\left(L+D-4\right)}{r^{2}}+\frac{\left(D-4\right)\left[\left(D-6\right)f+2rf^{\prime}\right]}{4r^{2}}\right)\Psi^{\text{V}}=0
\end{equation}
for the vector mode.

\subsection{Spin 2 Harmonic Decomposition}

The action for a spin 2 particle in a curved background is
\begin{equation}
S = -\frac{1}{2}\int \left(\nabla_\lambda h_{\mu\nu} \nabla^\lambda h^{\mu\nu} - 2\nabla_\lambda h_{\mu\nu} \nabla^{\nu} h^{\mu\lambda} + 2\nabla_\mu h \nabla_\nu h^{\mu\nu} - \nabla_\mu h \nabla^\mu h\right) \sqrt{-g} \dd[D]{x}.
\end{equation}
Working in the static limit, a harmonic decomposition of the metric perturbation into pieces with definite transformations
under $SO\left(D-1\right)$ again assists in isolating the physical degrees of freedom:
\begin{subequations}
\begin{align}
h_{tt}= & \sum_{\ell,M}f\left(r\right)H_{0}\left(r\right)Y_\ell^{M},\\
h_{tr}= & \sum_{\ell,M}H_{1}\left(r\right)Y_\ell^{M},\\
h_{rr}= & \sum_{\ell,M}f\left(r\right)^{-1}H_{2}\left(r\right)Y_\ell^{M},\\
h_{ti}= & \sum_{\ell,M}\left[\mathcal{H}_{0}\left(r\right)\nabla_{i}Y_\ell^{M}+h_{0}\left(r\right){Y_{i\hphantom{(T)}\ell}^{(T)M}}\right],\\
h_{ri}= & \sum_{\ell,M}\left[\mathcal{H}_{1}\left(r\right)\nabla_{i}Y_\ell^{M}+h_{1}\left(r\right){Y_{i\hphantom{(T)}\ell}^{(T)M}}\right],\\
h_{ij}= & \sum_{\ell,M}r^{2}\left[\mathcal{K}\left(r\right)\gamma_{ij}Y_\ell^{M}+G\left(r\right)\nabla_{(i}\nabla_{j)T}Y_\ell^{M} + h_2\left(r\right) \nabla_{(i} Y_{j) \hphantom{(T)}\ell}^{(T)\hphantom{\ell}M} + h_{T}\left(r\right)Y_{ij\ \ \ \ \ell}^{(TT)M}\right],
\end{align}
\end{subequations}
where the expression $\left(\dots\right)_{T}$ denotes the trace-free, symmetrized part of the enclosed indices, $\gamma_{ij}$ is the metric on the $\left(D-2\right)$-sphere and three types of hyperspherical harmonics have been used: transverse and traceless tensor harmonics $Y_{ij\ \ \ \ \ell}^{(TT)M}$, transverse vector harmonics $Y_{i\hphantom{(T)}\ell}^{(T)M}$ and scalar harmonics, $Y_{\ell}^{M}$. The spin 2 perturbation $h_{\mu\nu}$ is then naturally decomposed into scalar perturbations $H_{0}$, $H_{1}$, $H_{2}$, $\mathcal{H}_{0}$, $\mathcal{H}_{1}$, $G$ and $\mathcal{K}$, vector perturbations $h_{0}$, $h_{1}$ and $h_{2}$, and tensor perturbations $h_{T}$.

Gauge invariance is then exploited in order to work in the Regge-Wheeler gauge \cite{Regge:1957}, defined by the set of conditions
\begin{equation}
h_{2}=\mathcal{H}_{0}=\mathcal{K}=G=0.
\end{equation}
The tensor master variable $h_{T}$ decouples from all other modes, which is expressed through a field redefinition as
\begin{equation}
\Psi^{\text{T}}\equiv r^{\frac{D-2}{2}}h_{T}.
\end{equation}
The remaining vector perturbations can be expressed in terms of a single variable $Q$ through the equations of motion:
\begin{subequations}
\begin{align}
h_{0} & = -\frac{rf}{\left(\ell-1\right)\left(D-2+\ell\right)}\left[\left(D-2\right)Q+rQ^{\prime}\right],\\
h_{1} & = 0,
\end{align}
\end{subequations}
leading to the vector sector master variable
\begin{equation}
\Psi^{\text{RW}}\equiv\left(\frac{2r^{D-2}}{\left(\ell-1\right)\left(D-2+\ell\right)}\right)^{1/2}Q.
\end{equation}
For the scalar sector, $H_{1}$ is auxiliary and can be integrated out via its static equation of motion
\begin{equation}
H_{1} = 0.
\end{equation}
Further use of the equations of motion allow one to integrate
out auxiliary fields and arrive at an action for a single variable
\begin{equation}
\mathcal{V} = \mathcal{H}_{1}-\frac{\left(D-2\right)r}{2\ell\left(\ell+D-3\right)}H_{2},
\end{equation}
which is then field-redefined to
\begin{equation}
\Psi^{\text{Z}}\equiv\left(2fr^{D-4}\mathcal{F}\right)^{\frac{1}{2}}\mathcal{V},
\end{equation}
where
\begin{equation}
\mathcal{F}\equiv\frac{8\left(D-2\right)\ell\left(\ell+D-3\right)f\left[\left(D-3\right)\ell\left(\ell+D-3\right)-\left(D-3\right)\left(D-2\right)f-\left(D-2\right)rf^{\prime}\right]}{\left[2\ell\left(\ell+D-3\right)-2\left(D-2\right)f+\left(D-2\right)rf^{\prime}\right]^{2}}.
\end{equation}
The resulting equations of motion for the spin 2 master variables are
\begin{equation}
\label{eq:Spin2TensorEoM}
f\frac{\mathrm{d}^{2}\Psi^{\text{T}}}{\mathrm{d}r^{2}}+f^{\prime}\frac{\mathrm{d}\Psi^{\text{T}}}{\mathrm{d}r}-\left(\frac{\ell\left(\ell+D-3\right)+2\left(D-3\right)}{r^{2}}+f^{\prime}\frac{D-6}{2r}+f\frac{D\left(D-14\right)+32}{4r^{2}}\right)\Psi^{\text{T}}=0
\end{equation}
for the tensor modes, 
\begin{equation}
\label{eq:Spin2VectorEoM}
f\frac{\mathrm{d}^{2}\Psi^{\text{RW}}}{\mathrm{d}r^{2}}+f^{\prime}\frac{\mathrm{d}\Psi^{\text{RW}}}{\mathrm{d}r}-\left(\frac{\left(\ell+1\right)\left(D-4+\ell\right)}{r^{2}}+f\frac{\left(D-4\right)\left(D-6\right)}{4r^{2}}-f^{\prime}\frac{\left(D+2\right)}{2r}\right)\Psi^{\text{RW}}=0
\end{equation}
for the vector modes and
\begin{equation}
f\frac{\mathrm{d}^{2}\Psi^{\text{Z}}}{\mathrm{d}r^{2}}+f^{\prime}\frac{\mathrm{d}\Psi^{\text{Z}}}{\mathrm{d}r}-V_{\text{Z}}\left(r\right)\Psi^{\text{Z}}=0
\end{equation}
for the scalar modes, where
\begin{equation}
V_{\text{Z}}\left(r\right)=\frac{f\hat{V}_{\text{Z}}\left(r\right)}{4\left(D-2\right)r^{2}H\left(r\right)^{2}}
\end{equation}
and
\begin{subequations}
\begin{align}
H\left(r\right) &\equiv 2\ell\left(\ell+D-3\right)-2\left(D-2\right)f+\left(D-2\right)rf^{\prime},\\
& \hspace{-1cm} \begin{aligned}
\hat{V}_{\text{Z}}\left(r\right) & = 4\left(D-4\right)\left(D-2\right)^{4}f^{3}-8\left(D-2\right)^{2}\left[\left(D-2\right)\left(D-6\right)\ell\left(\ell+D-3\right)\right]f^{2}\\
 & \hspace{0.4cm} +4\left(D-2\right)\left[\left(D-2\right)\left(D-12\right)\ell^{2}\left(\ell+D-3\right)^{2}\right]f\\
 & \hspace{0.4cm}+2\left(D-2\right)^{3}\left(D+2\right)r^{3}f^{\prime3}-4\left(D-2\right)^{2}r^{2}\left[\left(D-6\right)\ell\left(\ell+D-3\right)\right]f^{\prime2}\\
 & \hspace{0.4cm}-8\left(D-2\right)^{2}\ell^{2}\left(\ell+D-3\right)^{2}rf^{\prime}+12\left(D-2\right)^{5}rf^{2}f^{\prime}\\
 & \hspace{0.4cm}+\left(D-2\right)^{3}\left(D\left(D+10\right)-32\right)r^{2}ff^{\prime2}\\
 & \hspace{0.4cm}-4\left(D-2\right)^{2}\left[\left(D-2\right)\left(3D-8\right)\ell\left(\ell+D-3\right)\right]rff^{\prime}\\
 & \hspace{0.4cm}+16L^{2}\left(\ell+D-3\right)^{2}\left[\left(D-2\right)\ell\left(\ell+D-3\right)\right].
 \end{aligned}
\end{align}
\end{subequations}

For every case considered in this paper except for the spin 2 scalar perturbations, the corresponding equations of motion have three regular singular points at the origin, the horizon and infinity and thus are described by hypergeometric equations. The equation of motion for spin 2 scalar perturbations, on the other hand, has four regular singular points. This is then described by a Heun function rather than a hypergeometric function. In the static limit however, it is possible to find an implicit field redefinition through a differential equation that maps the Heun equation of interest to a hypergeometric equation.\footnote{In $D=4$, this mapping becomes the familiar Chandrasekhar duality between the Zerilli and Regge-Wheeler variables \cite{Chandrasekhar:1985}; however, this duality does not exist in higher dimensions.}  For convenience of notation, the parameter $n\equiv D-2$ is now introduced. Setting $r_s=1$ for now, one finds

\begin{equation}
f\partial_r \left(f \partial_r Y^{\text{Z}}\right) - \frac{1}{4r^2}\Bigg[4 L\left(L + n-1\right)f - r^2 f'^2 \frac{2n - 1}{\left(n - 1\right)^2}+ 2rf' \frac{n}{n-1} + n\left(n - 2\right)\Bigg] Y^{\text{Z}} = 0,
\end{equation}
provided that $Y^{\text{Z}}$ satisfies
\begin{equation}
Y^{\text{Z}}=\frac{f^{1/2}}{rf^{\prime}}\left(\frac{rQ\left(r\right)\upsilon^{\prime}\left(r\right)}{4\left(m+\frac{1}{2}\left(n+1\right)nx\left(r\right)\right)}-\frac{P\left(r\right)\upsilon\left(r\right)}{16\left(m+\frac{1}{2}\left(n+1\right)nx\left(r\right)\right)^{2}}\right),
\end{equation}
where
\begin{equation}
m\equiv \ell\left(\ell+n-1\right)-n,
\end{equation}
the variable $\upsilon$ is related to $\mathcal{V}$ through
\begin{equation}
\upsilon\left(r\right)=\frac{\left(1-r^{1-n}\right)r^{\frac{n}{2}-1}}{2\left(L-1\right)\left(L+n\right)+n\left(n+1\right)r^{1-n}}\mathcal{V}\left(r\right),
\end{equation}
and
\begin{subequations}
\begin{align}
&\hspace{-0.9cm}\begin{aligned}
P\left(r\right) & \equiv 2\left(n^{2}-1\right)n\left(4m-n\left(n-2\right)\left(n+1\right)\right)x\left(r\right)^{2}\\
 & \hspace{0.4cm} +4m\left(n-1\right)n\left(3m+\left(n+1\right)n\right)x\left(r\right)+\left(n-1\right)\left(n+1\right)^{2}n^{3}x\left(r\right)^{3},
\end{aligned}\\
Q\left(r\right) & \equiv n\left(n-1\right)\left(n+1\right)x\left(r\right)^{2}-2\left(n-1\right)\left(m+\left(n+1\right)n\right)x\left(r\right),
\end{align}
\end{subequations}
with $x\left(r\right)\equiv\left(r_{s}/r\right)^{n-1}$. We finally match to the Zerilli variable used in \cite[eq. (4.18)]{Kol:2011} and define yet a new field
\begin{equation}
Z = \frac{1}{r^{3/2}\sqrt{1 - r^{d - 3}}} Y^{Z},
\end{equation}
for which the equation of motion becomes: 
\begin{equation}
\label{eq:Spin2ScalarEoM}
f\dv[2]{Z}{r} + \left(\frac{D - 6}{D - 3} f' + \frac{D}{r}\right)\dv{Z}{r} - \left(\frac{D - 4}{\left(D - 3\right)\left(D - 2\right)} f'' + \frac{\left(\ell - 1\right)\left(d + \ell - 2\right)}{r^2}\right) Z = 0.
\end{equation}
This is the starting point of our spin 2 scalar perturbations analysis in \autoref{ss:Spin2ScalarPerturbations}.

\section{Scalar in 5D Myers--Perry Harmonic Decomposition}
\label{a:MyersPerry}

As is the subject of \autoref{s:MyersPerry}, a ladder structure can also be found for scalar perturbations to a Myers--Perry black hole in 5D. In this appendix, we present the decomposition for a scalar in a 5D Myers--Perry background and re-derive its equation of motion following \cite{Rodriguez:2023}.

In Boyer-Lindquist coordinates, the Myers--Perry line element \cite{Myers:1986} is
\begin{equation}
\begin{aligned}\mathrm{d}s^{2}= & -\mathrm{d}t^{2}+\frac{\mu}{\Sigma}\left(\mathrm{d}t-a\sin^{2}\theta\mathrm{d}\phi-b\cos^{2}\theta\mathrm{d}\psi\right)^{2}+\frac{r^{2}\Sigma}{\Delta}\mathrm{d}r^{2}+\Sigma\mathrm{d}\theta^{2}\\
 & +\left(r^{2}+a^{2}\right)\sin^{2}\theta\mathrm{d}\phi^{2}+\left(r^{2}+b^{2}\right)\cos^{2}\theta\mathrm{d}\psi^{2},
\end{aligned}
\end{equation}
where
\begin{align}
\Sigma & \equiv r^{2}+a^{2}\cos^{2}\theta+b^{2}\sin^{2}\theta,\\
\Delta & \equiv\left(r^{2}+a^{2}\right)\left(r^{2}+b^{2}\right)-\mu r^{2}
\end{align}
and $0<r\leq\infty$; $0<\theta<\pi$; $0\leq\psi\leq2\pi$; and $0\leq\phi\leq2\pi$.
The three free parameters, $\mu$, $a$ and $b$ are
\begin{equation}
\label{eq:abDefinition}
\mu=\frac{8GM}{3\pi},\qquad a=\frac{3J_{\phi}}{2M},\qquad b=\frac{3J_{\psi}}{2M},
\end{equation}
where $M$ is the physical mass and $J_{\phi}$ and $J_{\psi}$ are
the conserved angular momenta corresponding to the Killing directions
$\phi$ and $\psi$ respectively. There are two horizons located at
the roots of $\left(r^{2}-r_{+}^{2}\right)\left(r^{2}-r_{-}^{2}\right)$
i.e. at
\begin{equation}
r_{\pm}^{2}=\frac{1}{2}\left(\mu-a^{2}-b^{2}\pm\sqrt{\left(\mu-a^{2}-b^{2}\right)^{2}-4a^{2}b^{2}}\right).
\end{equation}

The Klein-Gordon equation for a static scalar $\Psi$ in the 5D
Myers--Perry background is
\begin{equation}
\frac{1}{r}\partial_{r}\left(\frac{\left(r^{2}-r_{+}^{2}\right)\left(r^{2}-r_{-}^{2}\right)}{r}\partial_{r}\right)\Psi+M^{ij}\partial_{i}\partial_{j}\Psi+\nabla_{S^{3}}^{2}\Psi=0,
\end{equation}
where
\begin{equation}
\begin{aligned}
M^{ij}\partial_{i}\partial_{j} &= \frac{\left(b^{2}-a^{2}\right)\left(b^{2}+r^{2}\right)-b^{2}\mu}{\left(r^{2}-r_{+}^{2}\right)\left(r^{2}-r_{-}^{2}\right)}\partial_{\phi}^{2} + \frac{\left(a^{2}-b^{2}\right)\left(a^{2}+r^{2}\right)-a^{2}\mu}{\left(r^{2}-r_{+}^{2}\right)\left(r^{2}-r_{-}^{2}\right)}\partial_{\psi}^{2} \\
&\hspace{0.5cm} - 2\frac{ab\mu}{\left(r^{2}-r_{+}^{2}\right)\left(r^{2}-r_{-}^{2}\right)}\partial_{\phi}\partial_{\psi}
\end{aligned}
\end{equation}
and $\nabla_{S^{3}}^{2}$ is the Laplacian on the 3-sphere $\mathrm{d}\Omega_{3}^{2}=\mathrm{d}\theta^{2}+\sin^{2}\theta\mathrm{d}\phi^{2}+\cos^{2}\theta\mathrm{d}\psi^{2}$.
The equation admits an expansion in terms of hyperspherical harmonics,
\begin{equation}
\label{eq:MyersPerryHarmonicDecomp}
\Psi\left(x\right)=e^{i(m_{\phi}\phi+m_{\psi}\psi)}Y_{\ell}^{\ M}\left(\theta\right)R\left(r\right),
\end{equation}
where $Y_{\ell}^{\ M}\left(\theta\right)$ is an eigenfunction of
$\nabla_{S^{3}}^{2}$:
\begin{equation}
\nabla_{S^{3}}^{2}Y_{\ell}^{\ M}\left(\theta\right)=-\ell\left(\ell+2\right)Y_{\ell}^{\ M}\left(\theta\right).
\end{equation}
The radial equation then satisfies
\begin{equation}
\label{eq:RadialEq5SingPoints}
\frac{1}{r}\partial_{r}\left(\frac{\left(r^{2}-r_{+}^{2}\right)\left(r^{2}-r_{-}^{2}\right)}{r}\partial_{r}\right)R-\left(\ell\left(\ell+2\right)+M^{ij}\partial_{i}\partial_{j}\right)\Psi=0,
\end{equation}
where
\begin{equation}
m_{i}\mathrm{d}\theta^{i}=m_{\phi}\mathrm{d}\phi+m_{\psi}\mathrm{d}\psi.
\end{equation}

As is the case in Schwarzschild--Tangherlini, it is convenient to
define
\begin{equation}
\hat{\ell}\equiv\frac{\ell}{D-3}=\frac{\ell}{2}.
\end{equation}
The radial equation \autoref{eq:RadialEq5SingPoints} has five regular singular
points, but two of them are degenerate. We therefore define
\begin{equation}
\label{eq:xDefinition}
r^{2}=\frac{\left(r_{+}^{2}-r_{-}^{2}\right)x+r_{+}^{2}+r_{-}^{2}}{2}
\end{equation}
such that the inner and outer horizons are now located at $x=-1$
and $x=+1$ respectively, and the radial equation of motion becomes
\begin{equation}
\partial_{x}\left[\left(x^{2}-1\right)\partial_{x}R\right]-\left(\ell\left(\ell+2\right)+M^{ij}\partial_{i}\partial_{j}\right)R=0.
\end{equation}
Finally, we change the basis of $m_{i}$
to $\left(m_{L},m_{R}\right)$ defined by
\begin{align}
\label{eq:mLDefinition}
m_{\phi} & =m_{R}+m_{L},\\
m_{\psi} & =m_{R}-m_{L},
\end{align}
and then rescale each of these to
\begin{align}
\label{eq:mL}
\tilde{m}_{L} & =\frac{a-b}{r_{+}+r_{-}}\frac{m_{L}}{2},\\
\tilde{m}_{R} & =\frac{a+b}{r_{+}-r_{-}}\frac{m_{R}}{2}.
\label{eq:mR}
\end{align}
In terms of these the Klein-Gordon equation takes the form
\begin{equation}
\label{eq:EoM5DMyersPerry}
\partial_{x}\left[\left(x^{2}-1\right)\partial_{x}R\right]+2\left(\frac{\left(\tilde{m}_{L}^{2}+\tilde{m}_{R}^{2}\right)^{2}}{x-1}-\frac{\left(\tilde{m}_{L} -\tilde{m}_{R}\right)^{2}}{x+1}\right)R-\hat{\ell}\left(\hat{\ell}+1\right)R=0,
\end{equation}
which is the starting point of our analysis in \autoref{s:MyersPerry}.

\section{Deriving Ladders from Hypergeometric Identities}
\label{a:HypergeometricIdentities}

In each case considered in this paper, one finds that the equations of motion describing black hole perturbations in the static limit can be reduced to hypergeometric equations. One way to derive the ladder operators we use is therefore through the use of Gauss contiguous relations (although direct bootstrap methods can also be applied). These relate different hypergeometric functions with their parameters shifted by integer amounts.

The standard form of the hypergeometric equation is
\begin{equation}
\label{eq:Hypergeo}
x\left(1-x\right)u^{\prime\prime}\left(x\right)+\left[c-\left(a+b+1\right)x\right]u^{\prime}\left(x\right)-abu\left(x\right)=0,
\end{equation}
which has three regular singular points at $x=\left\{ 0,1,\infty\right\}$. This equation has two linearly independent solutions which, in the neighborhood of one of the singular points, take a simple form. The ladder operators derived in this paper act on \textit{any} solution of the equation of motion. A solution that is particularly convenient for deriving ladders is
\begin{equation}
\label{eq:HypergeometricFunction}
\left(-x\right)^{-a} F\left(a,1+a-c,1+a-b;\frac{1}{x}\right).
\end{equation}
since, in all of the cases of interest, the parameters $a$, $b$ and $c$ are of the form $a=\hat{\ell}+\text{constant}$, $b=\hat{\ell}+\text{constant}$, $c=2\hat{\ell}+\text{constant}$, such that the entries of this particular hypergeometric function solution becomes
\begin{equation}
F\left(\text{constant} + \hat{\ell}, \text{constant} - \hat{\ell}, \text{constant};\frac{1}{x}\right), 
\end{equation}
where all constants are independent of $\hat{\ell}$. The following identity will become useful: 
\begin{equation}
\label{eq:HypergeometricIdentity}
\begin{aligned}\left(x-1\right)\frac{\mathrm{d}}{\mathrm{d}x}F\left(\alpha,\beta,\gamma;\frac{1}{x}\right)= & -\frac{\alpha\left(\gamma-\alpha-1-\left(\beta-\alpha-1\right)\left(1-\frac{1}{x}\right)\right)}{\beta-\alpha-1}F\left(\alpha,\beta,\gamma;\frac{1}{x}\right)\\
 & +\frac{\alpha\left(\gamma-\beta\right)}{\beta-\alpha-1}F\left(\alpha+1,\beta-1,\gamma;\frac{1}{x}\right).
\end{aligned}
\end{equation}

Recalling the form of the solution of the equation of motion, \autoref{eq:HypergeometricFunction}, we want to turn \autoref{eq:HypergeometricIdentity}, an identity for the hypergeometric function alone,
into an identity for 
\begin{equation}
\label{eq:hyposol}
u\left(x\right)\equiv\left(-x\right)^{-\alpha}F\left(\alpha,\beta,\gamma;\frac{1}{x}\right).
\end{equation}
Taking a derivative gives
\begin{subequations}
\begin{align}
\frac{\mathrm{d}}{\mathrm{d}x}u\left(x\right) &= \frac{\mathrm{d}}{\mathrm{d}x}\left(\left(-x\right)^{-\alpha}F\left(\alpha,\beta,\gamma;\frac{1}{x}\right)\right), \\
&= \left(-x\right)^{-\alpha}\left[-\frac{\alpha}{x}F\left(\alpha,\beta,\gamma;\frac{1}{x}\right)+\frac{\mathrm{d}}{\mathrm{d}x}F\left(\alpha,\beta,\gamma;\frac{1}{x}\right)\right] \\
&= -\frac{\alpha}{x}u\left(x\right)+\left(-x\right)^{-\alpha}\frac{\mathrm{d}}{\mathrm{d}x}F\left(\alpha,\beta,\gamma;\frac{1}{x}\right),
\end{align}
\end{subequations}
thus
\begin{equation}
\frac{\mathrm{d}}{\mathrm{d}x}F\left(\alpha,\beta,\gamma;\frac{1}{x}\right)=\left(-x\right)^{\alpha}\left[\frac{\mathrm{d}}{\mathrm{d}x}u\left(x\right)+\frac{\alpha}{x}u\left(x\right)\right].
\end{equation}
Inserting into our identity \autoref{eq:HypergeometricIdentity}, we find
\begin{equation}
\begin{aligned}
\left(x-1\right)\left[\frac{\mathrm{d}}{\mathrm{d}x}u\left(x\right)+\frac{\alpha}{x}u\left(x\right)\right]= & -\frac{\alpha\left(\gamma-\alpha-1-\left(\beta-\alpha-1\right)\left(1- x^{-1}\right)\right)}{\beta-\alpha-1} u\left(x\right)\\
 & - \frac{\alpha\left(\gamma-\beta\right)}{\beta-\alpha-1} x u^{+}\left(x\right),
\end{aligned}
\end{equation}
where we have used $u^{+}\left(x\right)$ to denote the solution that will eventually correspond to $\hat{\ell}+1$: 
\begin{equation}
u^{+}\left(x\right)=\left(-x\right)^{-\alpha-1}F\left(\alpha+1,\beta-1,\gamma;\frac{1}{x}\right).
\end{equation}
Plugging in
\begin{equation}
\alpha = a, \qquad \beta = 1 + a - c, \qquad \text{and} \qquad \gamma = 1 + a - b,
\end{equation}
and isolating $u^+\left(x\right)$, we find
\begin{equation}
\label{eq:RaisingOperator}
u^{+}\left(x\right)=\frac{1}{x}\left[c\left(x-1\right)\frac{\mathrm{d}}{\mathrm{d}x}+ab\right]u\left(x\right).
\end{equation}
This differential operator on the right-hand side will gain the interpretation of being a raising operator, taking a solution at level $\hat{\ell}$ to level $\hat{\ell} + 1$. 

In order to find the lowering operator we can exploit the fact that hypergeometric functions are symmetric in the first two entries. This is useful because in swapping the first two entries we swap the location of $\hat{\ell}$ and $-\hat{\ell}$ in the hypergeometric function, so applying the identity \autoref{eq:HypergeometricIdentity}  again raises and lowers the opposite entry compared to before. Applying this to our swapped hypergeometric function therefore allows us to straightforwardly find the lowering operator. 

The key difference in this case is that, because the solution of the equation of motion is \autoref{eq:hyposol}, we must instead write this in terms of $\alpha$, $\beta$ and $\gamma$ as
\begin{equation}
u\left(x\right)\equiv\left(-x\right)^{-\beta}F\left(\alpha,\beta,\gamma;\frac{1}{x}\right).
\end{equation}
In other words, the power is now $\left(-x\right)^{-\beta}$ rather than $\left(-x\right)^{-\alpha}$, but the entries of the hypergeometric function are the same. Tracking this through the calculation means that taking a derivative now gives
\begin{equation}
\frac{\mathrm{d}}{\mathrm{d}x}F\left(\alpha,\beta,\gamma;\frac{1}{x}\right)=\left(-x\right)^{\beta}\left[\frac{\mathrm{d}}{\mathrm{d}x}u\left(x\right)+\frac{\beta}{x}u\left(x\right)\right]
\end{equation}
and we must now define $u^{-}\left(x\right)$, the solution that we will eventually find with $\hat{\ell}-1$, as
\begin{equation}
u^{-}\left(x\right)=\left(-x\right)^{-\beta+1}F\left(\alpha+1,\beta-1,\gamma;\frac{1}{x}\right).
\end{equation}
With these subtleties accounted for, we find
\begin{equation}
\begin{aligned}
\label{eq:LoweringOperator}
u^{-}\left(x\right) &= \Big[\left(c-2\right)x\left(x-1\right)\frac{\mathrm{d}}{\mathrm{d}x} \\
&\hspace{0.5cm} - \left(c-2\right)\left(c-1\right)+x\left(1-a-b-ab+c\left(a+b-1\right)\right)\Big]u\left(x\right).
\end{aligned}
\end{equation}
This differential operator will eventually be the lowering operator for the ladder in $\hat{\ell}$.

These raising and lowering operators take any solution of the equation of motion and return one with $\hat{\ell}$ raised or lowered by 1, provided the parameters $a$, $b$ and $c$ of the hypergeometric differential equation have the form $a=\hat{\ell}+\text{constant}$, $b=\hat{\ell}+\text{constant}$, $c=2\hat{\ell}+\text{constant}$. In order to find the ladders one therefore needs to first perform field redefinitions to reduce the equation of motion into the standard form in \autoref{eq:Hypergeo}, and then undo these field redefinitions on the raising and lowering operators, \autoref{eq:RaisingOperator} and \autoref{eq:LoweringOperator}, in order to find them in terms of the original variables appearing in the equations of motion.

\newpage


\bibliographystyle{jhep}
\bibliography{Bibliography.bib}

\providecommand{\href}[2]{#2}\begingroup\raggedright\begin{thebibliography}{10}

\bibitem{Charalambous:2021a}
P.~Charalambous, S.~Dubovsky and M.M.~Ivanov, \emph{{On the Vanishing of Love Numbers for Kerr Black Holes}}, \href{https://doi.org/10.1007/JHEP05(2021)038}{\emph{JHEP} {\bfseries 05} (2021) 038} [\href{https://arxiv.org/abs/2102.08917}{{\ttfamily 2102.08917}}].

\bibitem{Charalambous:2021b}
P.~Charalambous, S.~Dubovsky and M.M.~Ivanov, \emph{{Hidden Symmetry of Vanishing Love Numbers}}, \href{https://doi.org/10.1103/PhysRevLett.127.101101}{\emph{Phys. Rev. Lett.} {\bfseries 127} (2021) 101101} [\href{https://arxiv.org/abs/2103.01234}{{\ttfamily 2103.01234}}].

\bibitem{Charalambous:2022}
P.~Charalambous, S.~Dubovsky and M.M.~Ivanov, \emph{{Love symmetry}}, \href{https://doi.org/10.1007/JHEP10(2022)175}{\emph{JHEP} {\bfseries 10} (2022) 175} [\href{https://arxiv.org/abs/2209.02091}{{\ttfamily 2209.02091}}].

\bibitem{Hui:2022}
L.~Hui, A.~Joyce, R.~Penco, L.~Santoni and A.R.~Solomon, \emph{{Near-zone symmetries of Kerr black holes}}, \href{https://doi.org/10.1007/JHEP09(2022)049}{\emph{JHEP} {\bfseries 09} (2022) 049} [\href{https://arxiv.org/abs/2203.08832}{{\ttfamily 2203.08832}}].

\bibitem{Hui:2021}
L.~Hui, A.~Joyce, R.~Penco, L.~Santoni and A.R.~Solomon, \emph{{Ladder symmetries of black holes. Implications for love numbers and no-hair theorems}}, \href{https://doi.org/10.1088/1475-7516/2022/01/032}{\emph{JCAP} {\bfseries 01} (2022) 032} [\href{https://arxiv.org/abs/2105.01069}{{\ttfamily 2105.01069}}].

\bibitem{Berens:2022}
R.~Berens, L.~Hui and Z.~Sun, \emph{{Ladder symmetries of black holes and de Sitter space: love numbers and quasinormal modes}}, \href{https://doi.org/10.1088/1475-7516/2023/06/056}{\emph{JCAP} {\bfseries 06} (2023) 056} [\href{https://arxiv.org/abs/2212.09367}{{\ttfamily 2212.09367}}].

\bibitem{BenAchour:2022}
J.~Ben~Achour, E.R.~Livine, S.~Mukohyama and J.-P.~Uzan, \emph{{Hidden symmetry of the static response of black holes: applications to Love numbers}}, \href{https://doi.org/10.1007/JHEP07(2022)112}{\emph{JHEP} {\bfseries 07} (2022) 112} [\href{https://arxiv.org/abs/2202.12828}{{\ttfamily 2202.12828}}].

\bibitem{Sharma:2024}
C.~Sharma, R.~Ghosh and S.~Sarkar, \emph{{Exploring ladder symmetry and Love numbers for static and rotating black holes}}, \href{https://doi.org/10.1103/PhysRevD.109.L041505}{\emph{Phys. Rev. D} {\bfseries 109} (2024) L041505} [\href{https://arxiv.org/abs/2401.00703}{{\ttfamily 2401.00703}}].

\bibitem{Rai:2024}
M.~Rai and L.~Santoni, \emph{{Ladder symmetries and Love numbers of Reissner-Nordstr\"om black holes}}, \href{https://doi.org/10.1007/JHEP07(2024)098}{\emph{JHEP} {\bfseries 07} (2024) 098} [\href{https://arxiv.org/abs/2404.06544}{{\ttfamily 2404.06544}}].

\bibitem{Combaluzier-Szteinsznaider:2024sgb}
O.~Combaluzier-Szteinsznaider, L.~Hui, L.~Santoni, A.R.~Solomon and S.S.C.~Wong, \emph{{Symmetries of vanishing nonlinear Love numbers of Schwarzschild black holes}}, \href{https://doi.org/10.1007/JHEP03(2025)124}{\emph{JHEP} {\bfseries 03} (2025) 124} [\href{https://arxiv.org/abs/2410.10952}{{\ttfamily 2410.10952}}].

\bibitem{Kehagias:2024}
A.~Kehagias and A.~Riotto, \emph{{Black holes in a gravitational field: the non-linear static love number of Schwarzschild black holes vanishes}}, \href{https://doi.org/10.1088/1475-7516/2025/05/039}{\emph{JCAP} {\bfseries 05} (2025) 039} [\href{https://arxiv.org/abs/2410.11014}{{\ttfamily 2410.11014}}].

\bibitem{Gounis:2024}
L.R.~Gounis, A.~Kehagias and A.~Riotto, \emph{{The vanishing of the non-linear static love number of Kerr black holes and the role of symmetries}}, \href{https://doi.org/10.1088/1475-7516/2025/03/002}{\emph{JCAP} {\bfseries 03} (2025) 002} [\href{https://arxiv.org/abs/2412.08249}{{\ttfamily 2412.08249}}].

\bibitem{Berens:2025okm}
R.~Berens, L.~Hui, D.~McLoughlin, R.~Penco and J.~Staunton, \emph{{Geometric Symmetries for the Vanishing of the Black Hole Tidal Love Numbers}},  \href{https://arxiv.org/abs/2510.18952}{{\ttfamily 2510.18952}}.

\bibitem{Parra-Martinez:2025}
J.~Parra-Martinez and A.~Podo, \emph{{Naturalness of vanishing black-hole tides}},  \href{https://arxiv.org/abs/2510.20694}{{\ttfamily 2510.20694}}.

\bibitem{Lupsasca:2025}
A.~Lupsasca, \emph{{Why there is no Love in black holes}},  \href{https://arxiv.org/abs/2506.05298}{{\ttfamily 2506.05298}}.

\bibitem{Kol:2011}
B.~Kol and M.~Smolkin, \emph{{Black hole stereotyping: Induced gravito-static polarization}}, \href{https://doi.org/10.1007/JHEP02(2012)010}{\emph{JHEP} {\bfseries 02} (2012) 010} [\href{https://arxiv.org/abs/1110.3764}{{\ttfamily 1110.3764}}].

\bibitem{Cardoso:2018}
V.~Cardoso, M.~Kimura, A.~Maselli and L.~Senatore, \emph{{Black Holes in an Effective Field Theory Extension of General Relativity}}, \href{https://doi.org/10.1103/PhysRevLett.121.251105}{\emph{Phys. Rev. Lett.} {\bfseries 121} (2018) 251105} [\href{https://arxiv.org/abs/1808.08962}{{\ttfamily 1808.08962}}].

\bibitem{DeLuca:2022}
V.~De~Luca, J.~Khoury and S.S.C.~Wong, \emph{{Implications of the weak gravity conjecture for tidal Love numbers of black holes}}, \href{https://doi.org/10.1103/PhysRevD.108.044066}{\emph{Phys. Rev. D} {\bfseries 108} (2023) 044066} [\href{https://arxiv.org/abs/2211.14325}{{\ttfamily 2211.14325}}].

\bibitem{Barbosa:2025}
S.~Barbosa, P.~Brax, S.~Fichet and L.~de~Souza, \emph{{Running Love numbers and the Effective Field Theory of gravity}}, \href{https://doi.org/10.1088/1475-7516/2025/07/071}{\emph{JCAP} {\bfseries 07} (2025) 071} [\href{https://arxiv.org/abs/2501.18684}{{\ttfamily 2501.18684}}].

\bibitem{Caron-Huot:2025}
S.~Caron-Huot, M.~Correia, G.~Isabella and M.~Solon, \emph{{Gravitational Wave Scattering via the Born Series: Scalar Tidal Matching to $\mathcal{O}(G^7)$ and Beyond}},  \href{https://arxiv.org/abs/2503.13593}{{\ttfamily 2503.13593}}.

\bibitem{Gray:2024}
F.~Gray, C.~Keeler, D.~Kubiznak and V.~Martin, \emph{{Love symmetry in higher-dimensional rotating black hole spacetimes}}, \href{https://doi.org/10.1007/JHEP03(2025)036}{\emph{JHEP} {\bfseries 03} (2025) 036} [\href{https://arxiv.org/abs/2409.05964}{{\ttfamily 2409.05964}}].

\bibitem{Hui:2020}
L.~Hui, A.~Joyce, R.~Penco, L.~Santoni and A.R.~Solomon, \emph{{Static response and Love numbers of Schwarzschild black holes}}, \href{https://doi.org/10.1088/1475-7516/2021/04/052}{\emph{JCAP} {\bfseries 04} (2021) 052} [\href{https://arxiv.org/abs/2010.00593}{{\ttfamily 2010.00593}}].

\bibitem{Charalambous:2023}
P.~Charalambous and M.M.~Ivanov, \emph{{Scalar Love numbers and Love symmetries of 5-dimensional Myers-Perry black holes}}, \href{https://doi.org/10.1007/JHEP07(2023)222}{\emph{JHEP} {\bfseries 07} (2023) 222} [\href{https://arxiv.org/abs/2303.16036}{{\ttfamily 2303.16036}}].

\bibitem{Rodriguez:2023}
M.J.~Rodriguez, L.~Santoni, A.R.~Solomon and L.F.~Temoche, \emph{{Love numbers for rotating black holes in higher dimensions}}, \href{https://doi.org/10.1103/PhysRevD.108.084011}{\emph{Phys. Rev. D} {\bfseries 108} (2023) 084011} [\href{https://arxiv.org/abs/2304.03743}{{\ttfamily 2304.03743}}].

\bibitem{Glazer:2024}
D.~Glazer, A.~Joyce, M.J.~Rodriguez, L.~Santoni, A.R.~Solomon and L.F.~Temoche, \emph{{Higher-Dimensional Black Holes and Effective Field Theory}},  \href{https://arxiv.org/abs/2412.21090}{{\ttfamily 2412.21090}}.

\bibitem{Goldberger:2004jt}
W.D.~Goldberger and I.Z.~Rothstein, \emph{{An Effective field theory of gravity for extended objects}}, \href{https://doi.org/10.1103/PhysRevD.73.104029}{\emph{Phys. Rev. D} {\bfseries 73} (2006) 104029} [\href{https://arxiv.org/abs/hep-th/0409156}{{\ttfamily hep-th/0409156}}].

\bibitem{Tangherlini:1963}
F.R.~Tangherlini, \emph{{Schwarzschild field in n dimensions and the dimensionality of space problem}}, \href{https://doi.org/10.1007/BF02784569}{\emph{Nuovo Cim.} {\bfseries 27} (1963) 636}.

\bibitem{Solomon:2023ltn}
A.R.~Solomon, \emph{{Off-Shell Duality Invariance of Schwarzschild Perturbation Theory}}, \href{https://doi.org/10.3390/particles6040061}{\emph{Particles} {\bfseries 6} (2023) 943} [\href{https://arxiv.org/abs/2310.04502}{{\ttfamily 2310.04502}}].

\bibitem{Regge:1957}
T.~Regge and J.A.~Wheeler, \emph{{Stability of a Schwarzschild singularity}}, \href{https://doi.org/10.1103/PhysRev.108.1063}{\emph{Phys. Rev.} {\bfseries 108} (1957) 1063}.

\bibitem{PhysRevLett.24.737}
F.J.~Zerilli, \emph{Effective potential for even-parity regge-wheeler gravitational perturbation equations}, \href{https://doi.org/10.1103/PhysRevLett.24.737}{\emph{Phys. Rev. Lett.} {\bfseries 24} (1970) 737}.

\bibitem{Martel:2005}
K.~Martel and E.~Poisson, \emph{Gravitational perturbations of the schwarzschild spacetime: A practical covariant and gauge-invariant formalism}, \href{https://doi.org/10.1103/physrevd.71.104003}{\emph{Physical Review D} {\bfseries 71} (2005) }.

\bibitem{Lunin:2017}
O.~Lunin, \emph{{Maxwell{\textquoteright}s equations in the Myers-Perry geometry}}, \href{https://doi.org/10.1007/JHEP12(2017)138}{\emph{JHEP} {\bfseries 12} (2017) 138} [\href{https://arxiv.org/abs/1708.06766}{{\ttfamily 1708.06766}}].

\bibitem{Lunin:2025}
O.~Lunin, \emph{{Gravitational Waves in the Myers-Perry Geometry}},  \href{https://arxiv.org/abs/2510.14417}{{\ttfamily 2510.14417}}.

\bibitem{Myers:1986}
R.C.~Myers and M.J.~Perry, \emph{{Black Holes in Higher Dimensional Space-Times}}, \href{https://doi.org/10.1016/0003-4916(86)90186-7}{\emph{Annals Phys.} {\bfseries 172} (1986) 304}.

\bibitem{Chakraborty:2025}
S.~Chakraborty, P.~Heidmann and P.~Pani, \emph{{Fermionic Love of Black Holes in General Relativity}},  \href{https://arxiv.org/abs/2508.20155}{{\ttfamily 2508.20155}}.

\bibitem{Chandrasekhar:1985}
S.~Chandrasekhar, \emph{{The mathematical theory of black holes}}, Oxford University Press (1985).

\end{thebibliography}\endgroup

\end{document}